\documentclass[acmsmall,screen]{acmart}

\setcopyright{acmlicensed}
\copyrightyear{2023}
\acmYear{2023}
\acmDOI{10.1145/3593225}

\acmJournal{PACMHCI}
\acmVolume{7}
\acmNumber{EICS}
\acmArticle{173}
\acmMonth{6}
\acmPrice{15.00}

\usepackage{xcolor}
\usepackage{multirow}
\usepackage{makecell}
\usepackage{longtable}
\usepackage{float}

\newcommand{\refsection}[1]{\autoref{#1}}

\newcommand{\analysisPH}{A1--PH}
\newcommand{\analysisDC}{A2--DC}
\newcommand{\analysisDP}{A3--DP}
\newcommand{\analysisDE}{A4--DE}

\begin{document}

    \newcommand{\vzaffil}{%
        \institution{Technical University of Munich}
        \city{Munich}
        \country{Germany}%
    }

    \author{Patrik Zander}
    \affiliation{\vzaffil{}}
    \email{patrik.zander@tum.de}
    \orcid{0009-0000-2810-5558}

    \author{Valentin Zieglmeier}
    \affiliation{\vzaffil{}}
    \email{valentin.zieglmeier@tum.de}
    \orcid{0000-0002-3770-0321}

    \authorsaddresses{%
        Authors' addresses: Patrik Zander, patrik.zander@tum.de; Valentin Zieglmeier, valentin.zieglmeier@tum.de, Technical University of Munich, TUM School of Computation, Information and Technology, Chair of Software and Systems Engineering, Boltzmannstr. 3, 85748 Garching, Germany%
    }

    \title{Data Owner Benefit-Driven Design of People Analytics}
    \begin{abstract}
With increasingly digitalized workplaces, the potential for sophisticated analyses of employee data rises. This increases the relevance of people analytics (PA), which are tools for the behavioral analysis of employees. Despite this potential, the successful usage of PA is hindered by employee concerns. Especially in Europe, where the GDPR or equivalent laws apply, employee consent is required before data can be processed in PA. Therefore, PA can only provide relevant insights if employees are willing to share their data. One potential way of achieving this is the use of appeal strategies. In the design of PA, the core strategy that can be used is the inclusion of data owner benefits, such as automated feedback, that are given to employees in exchange for sharing their own data. In this paper, we examine benefits as an appeal strategy and develop four design principles for the inclusion of benefits in PA. Then, we describe an exemplary set of analyses and benefits, demonstrating how our principles may be put into practice. Based on this exemplary implementation, we describe and discuss the results of a user study ($n = 46$) among employees in the EU and UK. Our study investigates the factors that foster or hinder employees' consent to sharing their data with PA. Then, we introduce our data owner benefits and analyze whether they can positively influence this consent decision.

Our introduced data owner benefits were, contrary to our expectations, not suited to motivate our participants to consent to sharing their data. We therefore analyze how participants judge the benefits. Participants generally appreciate having them, confirming the value of including data owner benefits when designing PA. Some of our introduced benefits negatively influenced participants' sharing decision, though, meaning that careful consideration of potential risks is required when conceptualizing them.

    \end{abstract}

    %% Generate with http://dl.acm.org/ccs.cfm and copy here.
    % CAUTION! This MUST NOT be indented differently!


    % Ideally, does not contain anything from the abstract or title

    \maketitle

    \section{Introduction}
    \label{sec:introduction}

    The COVID-19 pandemic has accelerated the digitization of the modern workplace. As working spaces become
    increasingly digitized, more and more digital tools are used to manage and organize work.
    These tools collect data on various aspects of work, enabling more comprehensive analyses. This can be leveraged by people analytics (PA)---tools that analyze
    the data of individuals, typically employees, at the workplace~\cite{Tursunbayeva2018}. Their purpose is to provide
    various insights and aid the management of the company by allowing for a more objective and evidence-based
    way of making business decisions.

    Such analytics have been used to great effect in practice, showing promise in different areas of use~\cite{Davenport2010}.
    Despite this, there are also some barriers to the successful adoption of PA: For instance, the use of AI with historical data could
    cause biased decision-making if the data is not checked for biases~\cite{Kloepper2022}. Another
    important concern to mention is the privacy aspect: Due to their heavy reliance on data, PA tools require accurate information in order
    to function properly.

    Legal frameworks such as the General Data Protection Regulation (GDPR)~\cite{GDPR} require that employees consent before their data may be analyzed by PA.
    The disclosure of individuals depends on several factors, including, among others, the perceived risks and benefits of the disclosure~\cite{Smith2011}.
    This privacy calculus means that two approaches exist to increase employees' willingness to share: decreasing the risks, e.g. by providing additional transparency ~\cite{Zieglmeier2022a}, or increasing the appeal of the decision. To accomplish the latter,
    \citeauthor{Zieglmeier2022}~\cite{Zieglmeier2022} define three types of appeal strategies: values, incentives, and benefits.
    Unlike values, which guide the development of PA in an abstract manner and incentives, which are external to the tool, benefits
    are connected to the usage of the tool and therefore distributed during its usage~\cite{Zieglmeier2022}. As an example, consider
    a PA tool that measures the performance of employees -- one potential benefit that could be given to employees is automated
    feedback on where they could improve.

    The aim of this paper is to explore a shift in perspective for PA away from one that is predominantly focused on the
    needs of managers (who are generally the \textit{data consumers}~\cite{Pretschner2006} in our scenario) towards one that explicitly
    includes employee (\textit{data owner}~\cite{Pretschner2006}) benefits. Our work is structured as follows:
    First, we provide definitions for the fundamental concepts we are working with: our privacy model, people analytics,
    benefits, and the notions of \textit{data owner} and \textit{data consumer}. Then, we describe an abstract view of PA tools and
    propose a way data owner benefits could be incorporated into it.
    With the help of recommendations from existing literature, we conceptualize four principles for the data owner
    benefit-driven design of PA. Based on an exemplary implementation with a set of analyses and benefits, we demonstrate
    how our principles could be put into practice. To evaluate our principles, we then present the results
    of a user study we have conducted and explain potential threats to validity. Afterwards, we provide an insight into related
    literature and discuss our findings in the context of PA and existing design approaches.

    The results of our study reveal insights into the thought process of participants in a fictitious scenario based on our
    exemplary implementation. While participants typically recognized the value our analyses could provide, they also voiced concerns regarding
    the privacy implications of consenting to them. The more intrusive or risky the analysis was perceived to be,
    the fewer participants stated that they would consent to it. Contrary to our expectations, we find that the introduction of the benefits
    we conceptualized was not sufficient to significantly influence the willingness to disclose data. Consequently, we analyze the attitudes
    of our participants towards the benefits. We find that respondents generally appreciated them, with over 70\% of them preferring
    a scenario with benefits over a scenario without benefits. This highlights the value of introducing data owner benefits when
    designing and implementing PA. However, we also find that some participants feared negative consequences due to the benefits.
    This decreased their willingness to consent, which highlights that PA developers should think carefully about the potential
    risks of the benefits they introduce.

    \section{Theoretical foundations}
    \label{sec:foundation}

    We first introduce the concept of PA,
    as well as the terms data owner and data consumer. We then explain, using one particular privacy model~\cite{Smith2011}, how the decision of data owners can be influenced to encourage the sharing of data. Finally, we explain the concept of benefits as a concrete appeal strategy in the context of people analytics.

    \subsection{People analytics tools in the workplace}
    \label{subsec:people-analytics}

    The primary focus of this paper is on PA. These analyze personally identifiable information in the context
    of the workplace, most commonly for human resource management (HRM)~\cite{Tursunbayeva2018}.
    The goal of such tools is ``to help organisations understand their workforce as a whole,
    as departments or work groups, and as individuals, by making data about employee attributes,
    behaviour and performance more accessible, interpretable and actionable''~\cite[p. 224]{Tursunbayeva2018}.
    To achieve this, PA tools commonly utilize artificial intelligence to derive insights from employee data.

    While this opens up many opportunities for effective management~\cite{Huselid2018}, some authors have also
    pointed out risks of PA that may hinder adoption by companies~\cite{Kloepper2022,Tursunbayeva2022}.
    For instance, employees might not trust PA projects if they believe that their employers collect too much
    information about them, and thus be more reluctant to provide accurate information~\cite{Tursunbayeva2022}.

    Throughout this paper, we use the notions of \textit{data owner} and \textit{data consumer}~\cite[see][]{Pretschner2006} to refer to the people that are involved when a PA tool is used.
    Most commonly, employees represent the data owners and managers the data consumers.
    Depending on context and hierarchical structure, managers may also be data owners as well, though.

    \subsection{Privacy calculus and disclosure decisions}
    \label{subsec:privacy-model}

    As previously noted, perceived risks may hinder the willingness of a data owner to disclose accurate data.
    The privacy decisions made by individuals have been studied by several authors in various
    contexts~\cite[e.g.][]{Dijk2021, Phelps2000, Smith2011}.
    One notable example that explicitly models disclosure decisions is the APCO Macro Model~\cite[p.~998]{Smith2011}.
    It states that the decision to provide data depends on, among other factors, the privacy calculus, which itself depends
    on the perceived risks and benefits of the disclosure. Regarding PA, this indicates two possible ways of influencing the
    disclosure decision in favor of sharing data: By decreasing the perceived risks, e.g. by giving data owners greater transparency
    regarding how their data is used~\cite{Zieglmeier2022a} or by increasing the benefits of a disclosure,
    e.g. through the use of appeal strategies such as incentives~\cite{Zieglmeier2022}.

    \subsection{Benefits as an appeal strategy}
    \label{subsec:benefits}

    To increase the appeal of data disclosure, appeal strategies can be employed. \citeauthor{Zieglmeier2022}~\cite{Zieglmeier2022} define three
    dimensions:
    \textit{Values}, which are instilled into the PA tool at design time, tool-in\-de\-pen\-dent \textit{incentives},
    which managers can provide to employees when the PA system is already in use, and \textit{benefits}.
    Benefits are inherent affordances of a PA tool and
    ``inherent to the usage of the tool and irrevocably connected to it''~\cite[p. 219]{Zieglmeier2022}.
    They can range from an improved quality of information to personalized feedback for users~\cite{Zieglmeier2022}.

    \section{Approach}
    \label{sec:approach}

    In this section, we describe our approach in three steps:
    First, we introduce the fundamentals, assumptions, and conceptual limitations of our approach. Based on this, we propose a set
    of principles for the data owner benefit-driven development of PA tools, explaining their rationale and how they impact the
    development process of such a tool. Finally, we describe an implementation that incorporates our principles.

    \subsection{Introducing data owner benefits into PA}
    \label{subsec:approach-fundamentals}

    We start by presenting the status quo of PA, which focus on providing insights to data consumers.
    Then, we present the desired transformation and note conceptual limitations of our approach.

    \subsubsection{PA tools without data owner benefits}

    \begin{figure}[htbp]
      \includegraphics[width=.65\textwidth]{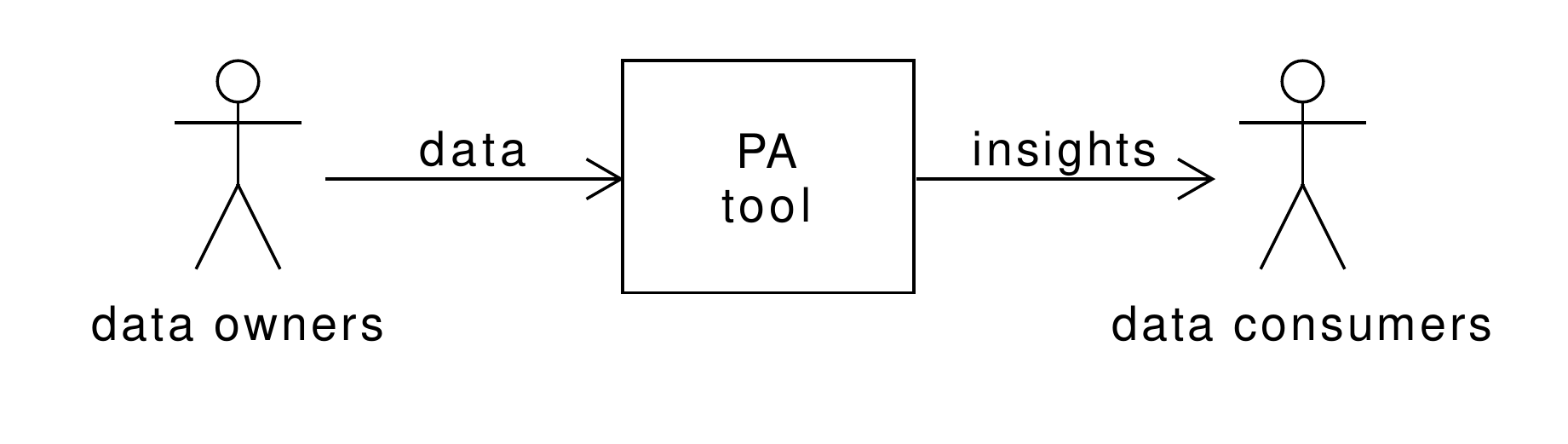}
      \caption{Informal model depicting the abstract perspective on a PA tool. The arrows symbolize the data flow between data owners, the tool itself, and data consumers.}
      \label{fig:abstract-pa-perspective}
      \Description{The model consists of three elements: the PA tool (represented as a box), data owners, and data consumers (both represented as stick figures). An arrow labeled ``data'' points from the data owners to the PA tool, and an arrow labeled ``insights'' points from the PA tool to the data consumers.}
    \end{figure}

    \autoref{fig:abstract-pa-perspective} shows an informal model of a high-level view on a PA tool:
    The tool collects personally identifiable data from data owners (usually employees) and, using some set of analyses, provides
    insights to data consumers (usually managers). This perspective places a large focus on the needs of data consumers. We assume that, in
    general, this perspective
    is the most prominent one in the development of PA tools. To justify this assumption, consider the software development
    process: As mentioned in \refsection{sec:foundation}, the primary goal of PA is to help organizations manage their
    workforce~\cite{Tursunbayeva2018}. As a result, the funding for PA projects is generally provided by the company,
    making it the client during the development process. Since software engineering requires
    ``the incremental evolution of the system toward a solution that is acceptable to the client''~\cite[p. 8]{Bruegge2009},
    this gives the company significant control over the requirements. If the company does not sufficiently take benefits of data
    owners into account (e.g. because it predominantly focuses on the management perspective), this may result in the view
    depicted in~\autoref{fig:abstract-pa-perspective}. Further indications can be found in real-world examples of PA tools:
    Human resource management tools such as \textit{Personio}\footnote{\url{https://www.personio.com/} (last accessed 2022-10-21)}
    are primarily marketed towards management~\cite{PersionioWebsite}, which aligns with the goal of such
    tools~\cite{Tursunbayeva2018}, but also indicates that there is a focus on the needs of data consumers.
    Another aspect that could be considered is the volume of information collected and processed by the PA tool.
    If large amounts of data are collected, data owners may be less likely to disclose accurate data~\cite{Tursunbayeva2018}
    -- at the same time, less collected data means fewer analyses for data consumers to use.
    As a real-world example, consider the reports provided by \textit{Microsoft Teams Analytics}~\cite{MicrosoftTeamsDocsReporting},
    which provide extensive insights ranging from the device usage~\cite{MicrosoftTeamsDocsDeviceUsageReport}
    to the activity of users based on their meetings, messages, and calls~\cite{MicrosoftTeamsDocsUserActivityReport}. We hypothesize
    that such analytics could be viewed as intrusive (see \refsection{subsec:results}). It should be noted that all of this is not to say that employees do not indirectly benefit from disclosing their data --
    it is just that the relationship between the disclosure and the positive effect may not be immediate, apparent or clear to the data
    owner in question.

    \subsubsection{Desired transformation} 
\begin{figure}
      \includegraphics[width=.65\textwidth]{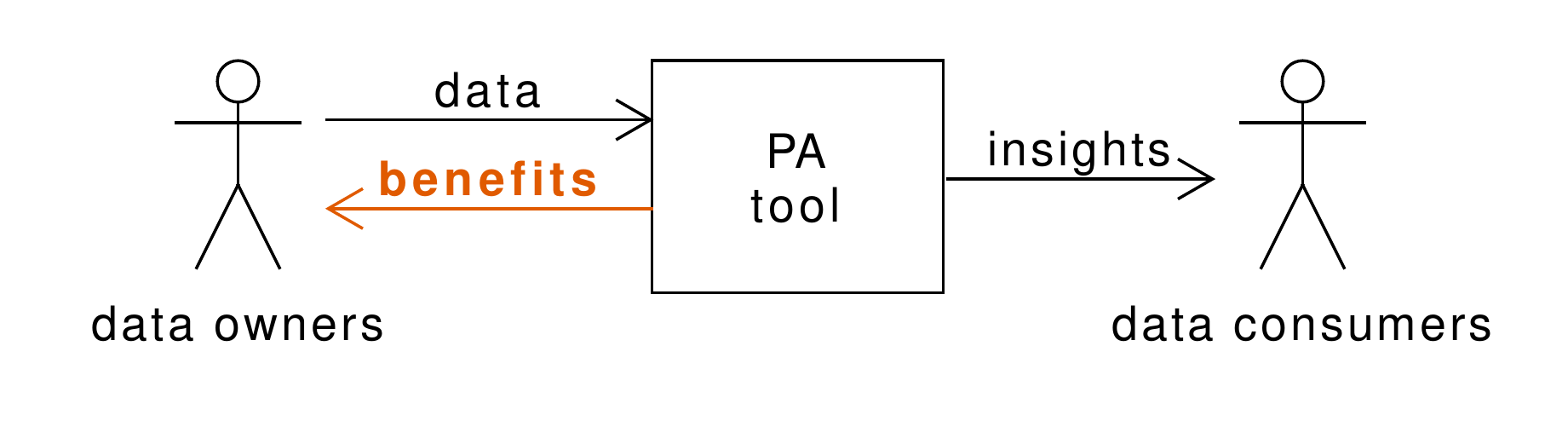}
      \caption{Informal model depicting a modified perspective on a PA tool, explicitly considering the benefits of data owners. The arrows symbolize the data flow between data owners, the tool itself, and data consumers.}
      \label{fig:abstract-pa-perspective-benefits}
      \Description{The model consists of three elements: the PA tool (represented as a box), data owners, and data consumers (both represented as stick figures). An arrow labeled ``data'' points from the data owners to the PA tool, and an arrow labeled ``insights'' points from the PA tool to the data consumers. Additionally, there is an orange arrow labeled ``benefits'' pointing back from the PA tool towards the data owners.}
    \end{figure}

\begin{figure}
      \includegraphics[width=\textwidth]{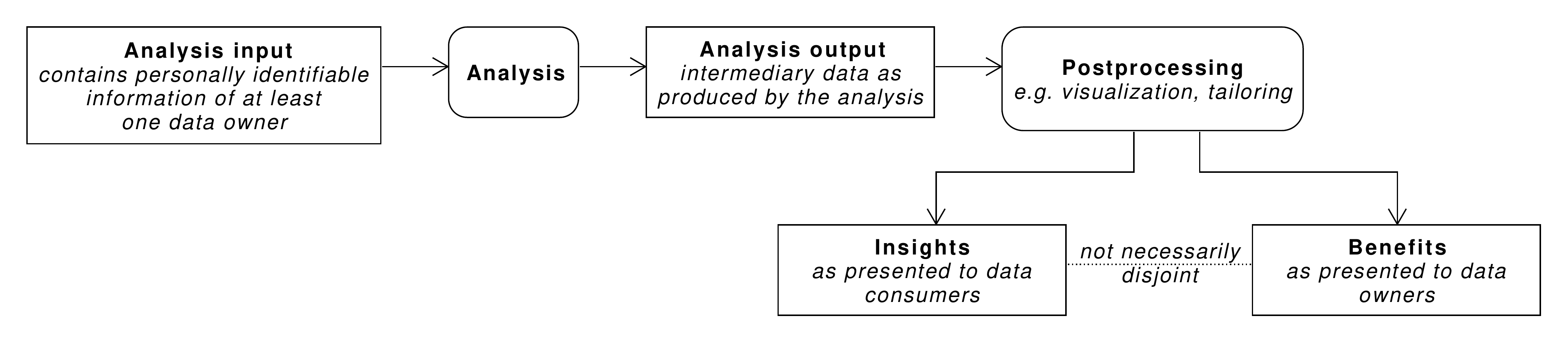}
      \caption{Informal model showing one potential way of modeling data owner benefits at the analysis level as discussed in \refsection{subsec:approach-fundamentals}. Regular rectangles represent data objects, serving as the input and output to processing steps, which are denoted using rounded rectangles.}
      \label{fig:pa-analysis-perspective}
      \Description{The model contains six nodes: ``Analysis input,'' which is noted to contain personally identifiable information of at least one data owner; ``Analysis''; ``Analysis output,'' labeled ``intermediary data as produced by the analysis''; ``Postprocessing,'' labeled ``e.g. visualization, tailoring''; ``Insights,'' labeled ``as presented to data consumers''; and ``Benefits,'' labeled ``as presented to data owners.'' Arrows connect the nodes as follows: from ``Analysis input'' to ``Analysis,'' from ``Analysis'' to ``Analysis output,'' and from ``Analysis output'' to ``Postprocessing.'' From ``Postprocessing,'' one arrow each goes to ``Benefits'' and ``Insights.'' Finally, the nodes ``Benefits'' and ``Insights'' are connected using a dotted line labeled ``not necessarily disjoint.''}
    \end{figure}

Our goal is to incorporate benefits into the design of PA tools as depicted in \autoref{fig:abstract-pa-perspective-benefits}. While there may be several ways of doing so, we have conceptualized one specific approach that ties benefits and insights to each individual analysis provided by the PA tool. This exemplary approach can be seen modeled in \autoref{fig:pa-analysis-perspective}: Data collected from data owners is processed by the analysis, which outputs an intermediate result. This intermediate result is then postprocessed in a separate step. During this postprocessing stage, no further analyses on the data should be conducted -- it should solely be used for purposes such as tailoring the results to a specific user's perspective or visualizing them. The result of postprocessing are a set of insights for data consumers and a set of benefits for data owners. Note that these results need not be disjoint -- in the simplest case, the insights gathered from the analysis could be forwarded as benefits to data owners.

    Modeling analyses, benefits, and insights this way provides multiple advantages that are relevant to our principles
    in \refsection{subsec:approach-principles}: For one, it requires developers to come up with data owner benefits as they
    conceptualize analyses for PA tools, which ensures that every analysis carries at least one data owner benefit that is
    inherently related to the insights generated by the analysis. In addition to this, the model does not prescribe a mechanism
    (e.g. dashboards, notifications) for the distribution of benefits and insights.

    \subsubsection{Conceptual limitations}

    It is important to note that the models in \autoref{fig:abstract-pa-perspective}, \autoref{fig:abstract-pa-perspective-benefits}, and \autoref{fig:pa-analysis-perspective} reveal three other aspects of PA that may influence its acceptance and usefulness within a company.
    These aspects go beyond the inclusion of benefits and can be seen as conceptual limitations of our approach. Consider the flow of data
    within a PA tool. First, the tool must collect the data, which poses the first challenge: the collected data must be available and suitable to derive the desired insight. If the data is insufficient or data owners refuse to disclose it, the analysis will not be useful. PA developers should carefully consider this when choosing the input data of each analysis. The next part is the actual analysis (i.e., the processing step), which has to transform the data. If the analysis is unable to deal with the input data or produce
    actionable insights, then its usefulness will be limited. Finally, even if the analysis is implemented perfectly, there is always
    a human component to consider. Ultimately, a manager has to interpret and act upon the generated insights. The manager may draw the wrong conclusions from the data and be unable to use PA to its
    full potential.

    \subsection{Principles of data owner benefit-driven design}
    \label{subsec:approach-principles}

    In the following, we conceptualize principles to aid the transformation described in the previous subsection.
    For each principle, we provide a description, a rationale as to why the principle makes sense,
    and how the principle may affect the development process. To further illustrate, we develop a running example
    that incorporates each of the principles.

    The principles were developed in a three-step process: They were initially prepared by the first author based on relevant
    literature. Then, the second author reviewed the principles and provided feedback, based on which the first author
    finalized the principles.
    In any case, the following principles are not to be understood as an exhaustive or universal list, as the development of PA,
    much like the development of all software, depends to a great extent on the requirements of the specific project~\cite{Bruegge2009a}.
    Furthermore, the conceptual limitations discussed in the previous subsection apply; our principles can only be considered applicable
    if these limitations are handled in suitable ways.

    \paragraph*{Running example} Consider an analysis that, based on the sentiment of messages, generates an insight about how
     the overall mood of a team of employees develops over time.

    \subsubsection{P1 -- Informed consent}

    Data owners must be informed about the purpose of analyses and given the choice to opt out.
    If a user chooses to opt out of certain analyses, the system must ensure that their data is not used for those analyses
    and revoke access to the corresponding benefits.

    \paragraph*{Rationale} Consider legislation such as the General Data Protection Regulation (GDPR) -- it requires that,
    before analyzing data, informed consent is retrieved from the owner of said data~\cite{GDPR}. With respect to privacy concerns,
    this informed consent can raise acceptance for analyses due to additional transparency~\cite{Kloepper2022,Tursunbayeva2018}.
    Given these points, data owners can be seen as \textit{data sovereigns} that are in charge of their data and must be convinced
    of its usage~\cite{Zieglmeier2022a}, which is where benefits~\cite{Zieglmeier2022} come into play.

    \paragraph*{Impact on development} The implementation must provide an opt-out page.
    Ideally, the analyses, including their insights and benefits, are explained there as well (see also P3).
    Additionally, the implementation must support the exclusion of data owners from analyses and benefits that have opted out.

    \paragraph*{Running example} The PA system provides an opt-out page where every analysis provided by the system
    is listed and explained. On this page, users can find the analysis introduced above, along with an explanation describing
    the data collected as well as the potential insights generated from the analysis.

    \subsubsection{P2 -- Conceptualization of analyses and benefits}
    \label{principles:p2}

    When conceptualizing the analyses and insights of a PA tool, the corresponding benefits (for data owners)
    should be devised as well. As mentioned in \refsection{subsec:approach-fundamentals}, each benefit could range from merely
    sharing the insights of the analysis with all involved data owners to providing a specialized, distinct view to each of them.

    \paragraph*{Rationale} Our hypothesis is that the inclusion of benefits will help increase the willingness of data owners to
    disclose their data for the purposes of PA. Furthermore, previous research suggests that it is important that
    ``employees experience the benefits of PA projects, and not just the organization''~\cite[p. 912]{Tursunbayeva2022}.

    \paragraph*{Impact on development} The most immediate consequence of this principle is that, during requirements elicitation,
    benefits need to be taken into account. This puts more emphasis on the needs of data owners throughout PA development,
    as their perspective must now explicitly be taken into account when the tool is constructed. Since changes to the
    requirements of a system tend to have significant impacts on other aspects of the system~\cite{Bruegge2009a}, we can derive further principles (see P3 and P4).

    \paragraph*{Running example} As a benefit, the insight could be provided without modification to the members of the team (i.e. the data owners).

    \subsubsection{P3 -- Distribution of benefits}

    The PA tool must be able to distribute benefits to data owners.
    This assumes appropriate access control to ensure that only the intended recipient of a benefit has access to it.

    \paragraph*{Rationale} This principle follows directly from the goal of this paper and the inclusion of benefits:
    Since we wish to make benefits accessible to data owners, a corresponding distribution mechanism must be part of the PA tool.

    \paragraph*{Impact on development}  Developers must account for this distribution mechanism during the design and implementation
    of the system. Such a mechanism may also include postprocessing (e.g. tailoring / visualization) as described
    in \refsection{subsec:approach-fundamentals}.

    \paragraph*{Running example} The analysis could be presented to managers as a graph on a dashboard.
    Since in this case, the benefits and insights are identical, the same graph can be presented to team members as well.

    \subsubsection{P4 -- Explanation and communication of benefits}

    Benefits must be explained in a clear and understandable way. This explanation must be accessible to data owners in a prominent location.

    \paragraph*{Rationale} If benefits are not perceived or understood, then data owners may not (accurately)
    take them into account when making disclosure decisions. Furthermore, transparency about the purpose of the collected data,
    which in our case includes the computation of data owner benefits, can promote trust in
    PA systems~\cite{Kloepper2022,Tursunbayeva2018}.

    \paragraph*{Impact on development} For each benefit, developers should devise a textual explanation describing
     what data owners receive if they consent to the corresponding analysis. This explanation could be checked
     against readability measures to ensure that it will be read and understood by data owners.\footnote{On a related note,
     \citeauthor{Fabian2017}~\cite{Fabian2017} have found that most privacy policies may be hard to read.}
     This explanation can be connected to the opt-out page described by P1: When consent is requested from the owner,
     a clear explanation of the gained benefits can be provided to allow an informed decision and positively influence
     the disclosure decision.

    \paragraph*{Running example} In the running example, the benefit is presented using a graph on a dashboard.
    The corresponding dashboard element could provide a tooltip describing that the graph depicts the mood of the team over
    time based on the sentiment expressed through comments sent by team members. This information is also presented on the opt-out page.

    \subsection{Exemplary implementation}

    \label{subsec:ex-impl}

    In the following, we demonstrate how our design principles can be applied to a realistic use case of PA.
    To make the analyses, their benefits, and the metrics more tangible, we first establish the context in which such a tool may be used.
    We then, based on existing literature and real-world examples, conceptualize a set of analyses and benefits, along with ways of visualizing and
    distributing them. To help potential users to quickly understand the usage and value proposition of the analyses, each analysis is intentionally simplified compared to related real-world examples.
    We close by describing the resulting proof of concept implementation and how our principles were realized. Note that the focus of
    this implementation is on the design principles, not on the concrete algorithms or correctness of the analyses themselves.

    \subsubsection{Context}
    According to \citeauthor{Rasmussen2015}~\cite{Rasmussen2015}, PA should always be devised with specific business
    challenges in mind. Such business challenges are dependent on the needs of the company seeking to utilize PA.
    In turn, these needs are dependent on the type of the business. We therefore narrow down the use case of our example.

    One potential use case of PA is in software development~\cite{Singer2017}. This scenario is useful for our case, as work in software development has been digitalized since the beginning, meaning that various mature tools exist that produce usable data for analyses.
    We model the structure of a fictitious software development company as follows:
    The company consists of teams of software developers, with each team working on a different software project.
    Each team has at least one manager. As is common in software development, various tools are used: Developers use a messenger such as Slack\footnote{\url{https://slack.com/} (last accessed 2022-10-21)} to communicate,
    an issue tracker (e.g., Jira\footnote{\url{https://www.atlassian.com/software/jira} (last accessed 2022-10-21)}) to manage their tasks, and a version control system such as
    Git\footnote{\url{https://git-scm.com/} (last accessed 2022-10-21)} to keep track of changes to the source code of their projects. The data collected by these tools can be used to conduct analyses, which will be explained in the following.

    \subsubsection{Analyses and benefits} Having established the context for which our PA tool is intended, we devise a set of analyses.
    Each analysis is accompanied by a description indicating the business challenge it aims to address, the data sources of the analysis,
    and how the analysis can be visualized. In accordance with \hyperref[principles:p2]{P2}~(Conceptualization of analyses and benefits),
    we define data owner benefits along with the analyses.

\begin{figure}[htbp]
      \includegraphics[width=.6\textwidth]{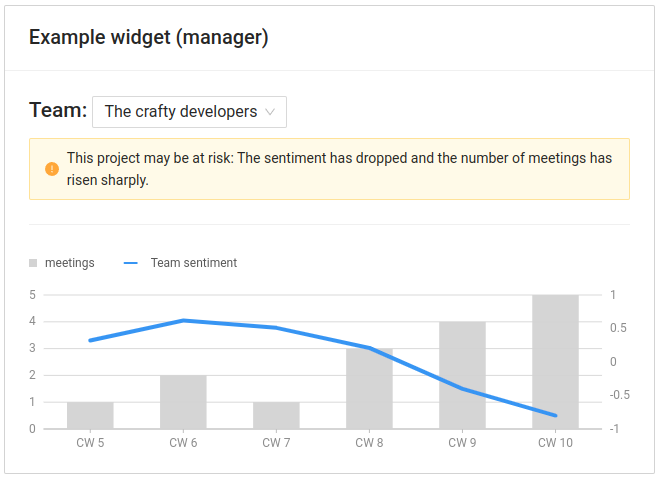}
      \caption{Dashboard widget related to the analysis \analysisPH. The bars of the graph indicate the number of meetings per calendar week, while the line indicates the average sentiment expressed in messages of developers in that week. Higher values indicate an overall better sentiment.}
      \label{fig:a1}
      \Description{Depicted is a dashboard widget titled ``Example widget (manager).'' Beneath the title, it offers a dropdown menu for selecting a team, below which there is a yellow alert message. It reads, ``This project may be at risk: The sentiment has dropped and the number of meetings has risen sharply.'' Below the alert is a combined bar (for the number of meetings) and line graph (for the team's sentiment value). The x-axis is labeled from calendar week 5 to calendar week 10, in increments of one. The development of the graph matches the alert: The team sentiment gradually decreases towards calendar week 10, and the number of meetings reaches its highest point during that week after a sudden increase in calendar week 8.}
    \end{figure}

    \paragraph{\analysisPH: Are there any projects at risk of failure? (project health)}
    This analysis could rely on data such as sentiment of messages written by developers, as well as changes in the meeting
    profiles of teams: A drop in sentiment or a sudden increase in the number of meetings may indicate that a project
    is at risk~\cite{Singer2017}. One example of a tool with the potential of analyzing meeting-related statistics is Microsoft Teams, which can track the number of meetings of a user~\cite{MicrosoftTeamsDocsUserActivityReport}. For sentiment analysis, an example is the PA software by Jive Software~\cite{JiveSoftware}, which can analyze sentiment within teams to gauge employee satisfaction. Our example analysis can be visualized using a graph as
    depicted in \autoref{fig:a1}.
    If a data owner consents to the use of their data for the sake of this analysis,
    they receive a similar dashboard widget, limited to their team's data and including additional information
    about their own sentiment compared to the group. If they do not, they are excluded from the calculation; the sentiment of their
    messages is not used when computing the team's average sentiment.

\begin{figure}[htbp]
      \includegraphics[width=.65\textwidth]{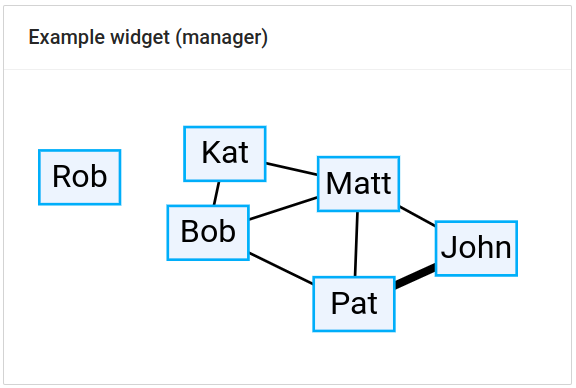}
      \caption{Dashboard widget related to the analysis \analysisDC.
      The connections between the nodes of the graph indicate which developers interact with each other.}
      \label{fig:a2}
      \Description{Depicted is a dashboard widget titled ``Example widget (manager).'' Beneath the title, there is a graph with several nodes representing various developers. Collaboration is represented using undirected edges between the nodes. The node labeled ``Rob'' appears to be isolated from all other nodes. Compared to edges between other nodes (e.g. ``Kat'' and ``Bob''), the connection between ``John'' and ``Pat'' is thicker.}
    \end{figure}

    \paragraph{\analysisDC: Which developers interact with each other? (developer collaboration)}
    As mentioned before, the communication between developers may be used as a data source for analyses.
    \citeauthor{Singer2017}~\cite{Singer2017} give an overview of different ways that interactions between developers can be leveraged.
    As an example, knowing which developers interact with each other can help detect isolated developers
    and allow managers to intervene accordingly~\cite{Singer2017}. Alternatively, one could examine how strongly
    interconnected each team is to gain an insight into how well its developers work together as proposed
    by \citeauthor{Leonardi2018}~\cite{Leonardi2018}.
    In practice, Polinode~\cite{Pitts2015} is one such example.
    This tool visualizes in a graph which employees commonly work together.
    \autoref{fig:a2}, then, shows our implemented dashboard widget depicting such an
    analysis.
    As benefit, a data owner may receive a dashboard widget with an individualized graph -- one that shows
    their connections to other people within the company. Should the data owner refuse to provide their data, they will not
    receive such a widget, their messages will not be processed, and they will not be included as a node in any graph pertaining to this analysis.

\begin{figure}[htbp]
      \includegraphics[width=.7\textwidth]{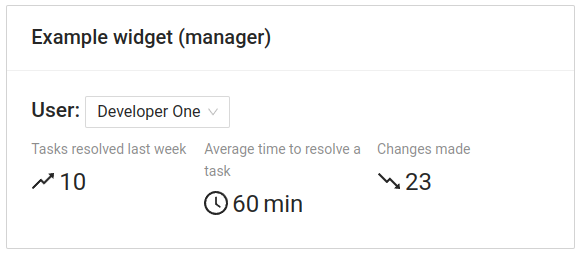}

      \caption{Dashboard widget related to the analysis \analysisDP. It depicts three activity indicators for each developer: The number of tasks resolved, the average time done to resolve a task, and the number of commits on the main branch of the code repository.}
      \label{fig:a3}
      \Description{Depicted is a dashboard widget titled ``Example widget (manager).'' Beneath the title, it offers a dropdown menu for selecting a user. Beneath the dropdown menu, the widget displays three statistics in a three-column layout: ``Tasks resolved last week,'' ``Average time to resolve a task,'' and ``Changes made.'' In this example, the selected user ``Developer One'' has resolved 10 tasks last week, taken 60 minutes on average to resolve a task, and made 23 changes. Through upwards- and downwards-trending arrows, the widget indicates that the number of resolved tasks has increased, whereas the number of changes made has gone down.}
    \end{figure}

    \paragraph{\analysisDP: Which of our developers are performing well, who may need support? (developer performance)}
    Another potential use of people analytics is to examine the performance or productivity of employees.
    \citeauthor{Davenport2010}~\cite{Davenport2010} name an example where Google used analytics to track the employees with the
    highest and the lowest performance. In software development, information indicating the performance of employees could, for instance,
    be derived from issue tracking data: A simple metric for determining the productivity of a developer may be based on how many
    issues they resolve in a given amount of time or how many contributions they have made. Indeed, developer analytics such as IBM Cloud DevOps Insights~\cite{CroninIBM} leverage information from version control and issue trackers, demonstrating that these data sources can be valuable in practice.
    \autoref{fig:a3}, accordingly, shows our widget displaying performance statistics of developers.
    As for benefits, consenting developers
    may receive the same statistics about themselves, along with periodical feedback on how these statistics develop over time.
    In case the data owner does not consent, their data will not be used for feedback or performance statistics. In this case,
    the dashboard widget will not allow its users to select the corresponding data owner.

\begin{figure}[htbp]
      \includegraphics[width=.7\textwidth]{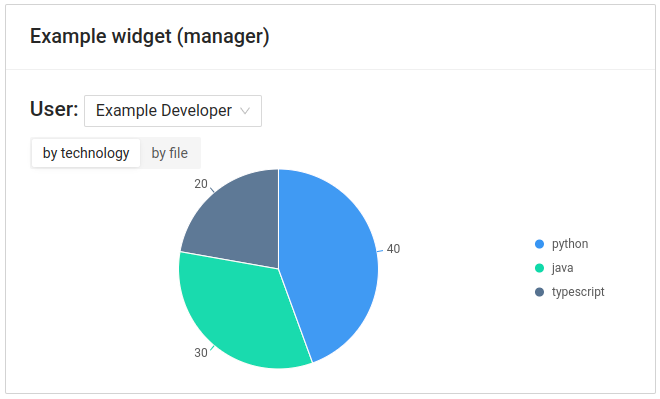}

      \caption{Dashboard widget related to the analysis \analysisDE. The slices of the pie chart represent the share of each technology in the developer's work as tracked by version control.}
      \label{fig:a4}
      \Description{Depicted is a dashboard widget titled ``Example widget (manager).'' Beneath the title, it offers a dropdown menu for selecting a user. Below this menu is a toggle offering the options ``by technology'' and ``by file,'' with the former being selected. Underneath, there is a pie chart with three slices: typescript (at 20), java (at 30), and python (at 40).}
    \end{figure}

    \paragraph{\analysisDE: Which technologies does each developer have the most experience with? (developer expertise)}
    Finally, version control information can be used to examine what technologies (and files) a developer is most accustomed to,
    which can aid managers when finding suitable developers for a task. As \citeauthor{Guzzi2012}~\cite{Guzzi2012} demonstrate,
    such a metric can be used to aid in finding the right developer to ask questions about a certain file.
    The corresponding dashboard widget is shown in \autoref{fig:a4}.
    Since this analysis, just like \analysisDP{}, draws from version control data, the IBM Cloud Dev\-Ops Insights~\cite{CroninIBM} suite can be named as a real-world instance of such an analysis. As with \analysisPH{} and \analysisDC, the corresponding benefit can be realized by providing data owners with the same insight, tailored to their own data.
    If a data owner does not consent to this analysis, this analysis behaves analogously to \analysisDP: the data is not processed, and the data owner
    cannot be selected in the widget under any circumstances.

    \subsubsection{Implementation}

    Our proof of concept implementation offers two variants of an opt-out page:
    One that only presents the analyses without noting any data owner benefits,
    and one that explicitly links consent to potential benefits received. Each description provides information about the collected
    data and the intended insight of the analysis. Users are further informed about the consequences of consenting:
    They receive information about what the manager will see (i.e., the previously presented dashboard widgets) and,
    in case benefits are enabled, the description is extended by information about the provided benefits.
    To make sure the benefits would be perceived during the evaluation, the changes were highlighted in yellow.
    The concrete descriptions for each analysis can be found
    in \hyperref[appendix:analysis-descriptions]{Appendix C}.

    The aforementioned features, analyses, and benefits reflect our principles:
    P1 (Informed consent) and P4 (Explanation and communication of benefits) are realized through the textual descriptions on the opt-out
    pages, and the exemplary dashboard widgets represent P3 (Distribution of benefits). Additionally, P2 (Conceptualization of analyses
    and benefits) is realized, as we conceptualized all analyses together with their benefits.

    \section{Evaluation}

    To assess the impact of benefits on participants' willingness to disclose their data, we conduct a mixed-method evaluation.
    This section is dedicated towards describing the design of our study, formulating corresponding research questions, and
    describing our findings.

    \subsection{Method}
    \label{subsec:eval-method}

    The study was conducted using an eight-page questionnaire. In this subsection, we describe the design of our study, the questionnaire
    our participants answered, and the methods we used to analyze our results.

    \subsubsection{Questionnaire}

    Our questionnaire consists of eight pages of questions and one welcome page with a privacy statement and followed a before / after
    design. To ensure compliance with the GDPR, potential participants were first informed about the purpose of the study and how their
    data would be processed. Those that did not consent to the use of their data were asked to return their submissions on Prolific;
    their data was not collected.

    At the beginning of the survey, participants were asked about their age (Q1), gender (Q2), employment status (Q3),
    whether they would classify their job as closer to that of a manager or an employee (Q4), and if their job is closer
    to a white-collar job or a blue-collar job (Q5). Participants were then presented with a scenario: They would assume the
    role of a software developer in a fictitious software development company. The scenario description was accompanied by a
    brief explanation of Jira, Slack, Git, and the concept of PA.

    Participants were then shown the privacy settings depicted in \hyperref[appendix:analysis-descriptions]{Appendix C} without benefits,
    asking for each analysis whether they would consent in the given scenario (Q6 -- Q9 for \analysisPH~to \analysisDE). To ensure that
    even non-technical respondents could follow the descriptions, we minimized the use of technical terms in favor of simpler language.
    Afterwards, we asked participants whether they felt like they understood the analysis descriptions (Q10) and to mention anything that
    was unclear (Q11).  After an attention check (Q12), respondents were then prompted to give an insight into their decision-making process
    for each analysis (Q13 -- Q16).

    Once participants had completed the first part, they were given a description of a change in the scenario:
    The company would now introduce data-owner benefits distributed to each employee. Afterwards, we showed them a second version of the privacy
    settings page, this time including the data owner benefits conceptualized in \refsection{subsec:ex-impl}.
    To ensure that the benefits would be perceived, we highlighted all changes with a yellow background color (see \hyperref[appendix:analysis-descriptions]{Appendix C}).
    Once more, participants were asked whether they would or would not consent to each analysis (Q17 -- Q20),
    given the modified scenario. Similarly to before, we then asked if participants understood the benefits (Q21 and Q23),
    if anything was unclear (Q22), and to explain their reasoning once more, this time taking into account the benefits and their
    impact on the decision (Q24 -- Q27). Finally, participants were asked which variant of the scenario
    they preferred (Q28), another attention check (Q29), and whether they had any additional remarks (Q30).
    The concrete questions of our questionnaire can be found in \hyperref[appendix:questionnaire]{Appendix A}.

    As a result of this design, all participants completed the questionnaire in the same order. We intentionally chose this design, as we consider this the more realistic scenario in the real world. Employees will usually be faced with an existing PA tool that they are used to working with, and be newly introduced to benefits rather than losing them. Furthermore, as we focus on the qualitative analysis, we decided against a control group, as that would have reduced the number of participants exposed to our intervention.

    \subsubsection{Data collection}
     We recruited a total of 51 participants (not including six users who canceled their submission),
     one of which was rejected, via Prolific.\footnote{\url{https://www.prolific.co/} (last accessed 2022-10-21)}
     To narrow down the sample of eligible users, we utilized Prolific's demographic prescreening.\footnote{\url{https://researcher-help.prolific.co/hc/en-gb/articles/360009221093-How-do-I-use-Prolific-s-demographic-prescreening-} (last accessed 2022-10-21)}
     We selected according to three criteria:

     \begin{itemize}
      \item \textbf{Fluent language}: English, since that is the language of our questionnaire
      \item \textbf{Employment status}: Part-time or full-time, to recruit users that are familiar with the workplace setting, thus
      increasing the likelihood of realistic answers
      \item \textbf{Country of residence}: United Kingdom or any country within the European Union, to ensure that all participants
      reside in regions that fall under the GDPR~\cite{GDPR} or GDPR-equivalent legislation.\footnote{https://www.gov.uk/data-protection (last accessed 2022-10-21)}
      This is because we want to ensure a common frame of regulation among all participants -- after all, our privacy model~\cite{Smith2011} includes
      a relationship between the privacy concerns of an individual and regulation, meaning that privacy issues could be viewed
      differently depending on what laws are in effect.
     \end{itemize}

    \subsubsection{Data analysis}

    To analyze the results, we used both qualitative and quantitative methods as described below.

    \paragraph{Qualitative analysis} In order to determine the factors that influenced the privacy choices of our participants
    and to assess the impact of data owner benefits on their decisions, we analyzed the answers to the questions Q13 -- Q16 and
    Q24 -- Q27 using the following process: In a first step, both authors worked on the first half
    of the data independently -- they then met and agreed upon a common coding scheme for both sets of questions. Afterwards,
    both authors coded all of the answers using this new scheme, adding new codes where appropriate. The assigned codes were then
    finalized by the first author. The resulting coding scheme along with examples can be found in
    \hyperref[appendix:codes]{Appendix B}. The coding itself was inspired by recommendations of \citeauthor{Adu2021}~\cite{Adu2021}.

    \paragraph{Quantitative analysis} In addition to the qualitative analysis of the free-text answers mentioned above, we performed
    McNemar's test~\cite{McNemar1947} using the exact test (described in e.g. \cite{Fay2010}) to assess whether there
    was a significant impact of the introduced benefits on the willingness of our participants to consent to each analysis. This
    was made possible by the before / after design of our study.

    \subsection{Research questions}

    We examine the following research questions:

    \begin{itemize}
      \item[RQ1:] Did participants understand the analysis and benefit descriptions provided in the survey?
      \item[RQ2:] How did respondents view each analysis, which factors influenced their initial privacy decisions?
      \item[RQ3:] Did data owner benefits significantly influence the outcome of privacy decisions in our scenario?
      \item[RQ4:] How did participants perceive the introduced data owner benefits?
      \item[RQ5:] Overall, did participants prefer the introduced people analytics with or without data owner benefits?
    \end{itemize}

    \subsection{Results}
    \label{subsec:results}
    In total, 57 people started the survey, 51 of which successfully completed it.
    One participant was excluded due to low-effort answers as per Prolific's rejection policy.\footnote{\url{https://researcher-help.prolific.co/hc/en-gb/articles/360009092394-Approvals-rejections-returns} (last accessed 2022-10-19)}
    Of the remaining 50 participants, an additional four were excluded due to failing an attention check
    (i.e. providing an incorrect answer to Q12 or Q29) or if their Likert scale responses to Q21 and Q23 differed by at
    least two points (e.g. "Strongly agree" and "Neither agree nor disagree"). This left us with 46 viable sets of answers which are
    described in the following. Each respondent has been given an identifier, R1, which we will use to refer to answers written by them.

    \subsubsection{Demographics} The demographic breakdown of our answers is shown in \autoref{tab:demographics}. We observe
    that our sample primarily contains people under the age of 45 (about 87\%) and more men than women. Almost a quarter (23.9\%)
    of respondents indicated that their role in the workplace was closer to that of a manager than to that of an employee.

    \begin{table}[htbp]
      \caption{The demographics of our survey. All percentages have been rounded to one decimal place.}
      \begin{tabular}{llll}
        \toprule
        \textbf{Question} & \textbf{Option} & \textbf{Number} & \textbf{Percentage} \\
        \midrule
        Q1 (age)                        & 24 or younger                                    & 17     & 37.0\%                  \\
                                        & 25 -- 34                                          & 18     & 39.1\%                 \\
                                        & 35 -- 44                                          & 5      & 10.9\%                 \\
                                        & 45 -- 54                                          & 3      & 6.5\%                  \\
                                        & 55 -- 64                                          & 3      & 6.5\%                  \\
                                        & 65 or older                                       & 0      & 0.0\%                  \\
        \midrule
        Q2 (gender)                      & Male                                             & 28     & 60.9\%                   \\
                                        & Female                                           & 18     & 39.1\%                   \\
                                        & Other                                            & 0      & 0.0\%                    \\
                                        & Prefer not to say                                & 0      & 0.0\%                    \\
        \midrule
        Q3 (employment status)           & Unemployed                                       & 0      & 0.0\%                    \\
                                        & Student                                          & 10     & 21.7\%                   \\
                                        & Employed                                         & 32     & 69.6\%                   \\
                                        & Self-employed                                    & 4      & 8.7\%                    \\
                                        & Retired                                          & 0      & 0.0\%                    \\
                                        & Other                                            & 0      & 0.0\%                    \\
        \midrule
        Q4 (role in employment)          & Manager                                          & 11     & 23.9\%                   \\
                                        & Employee                                         & 31     & 67.4\%                   \\
                                        & Unsure                                           & 1      & 2.2\%                    \\
                                        & Not applicable                                   & 3      & 6.5\%                    \\
        \midrule
        Q5 (white or blue collar)        & White-collar                                     & 33     & 71.7\%                   \\
                                        & Blue-collar                                      & 7      & 15.2\%                   \\
                                        & Unsure                                           & 3      & 6.5\%                    \\
                                        & Not applicable                                   & 3      & 6.5\%                    \\
\bottomrule
        \end{tabular}
        \label{tab:demographics}
    \end{table}

    \subsubsection{RQ1 -- Did participants understand the analysis and benefit descriptions provided in the survey?}
\begin{figure}[htbp]
        \includegraphics[width=\textwidth]{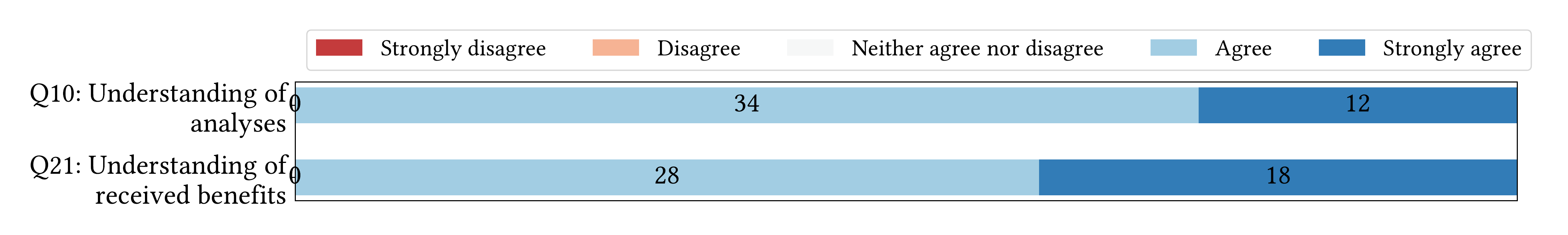}
        \caption{Answers to questions Q10 (agreement with the statement 'I felt like I understood all analyses (A1 to A4).") and Q21 (agreement with the statement 'I felt like I understood what benefits I would receive if I consented to an analysis.'),
    		both on a Likert scale.}
        \label{fig:rq1-understanding}
        \Description{Depicted are two horizontally-stacked bar graphs representing the agreement to questions Q10 and Q21. The possible responses were ``Strongly disagree,'' ``Disagree,'' ``Neither agree nor disagree,'' ``Agree,'' and ``Strongly agree''. For Q10, labeled ``Understanding of analyses,'' the responses were 12 times ``Strongly agree'' and 34 times ``Agree.'' For Q21, labeled ``Understanding of received benefits,'' the responses were 18 times ``Strongly agree'' and 28 times ``Agree.'' No participant chose any of the other response options for either question.}
    \end{figure}

    As mentioned in \refsection{subsec:approach-principles}, one of our design principles specifies that data owners must be able to
    reach an informed decision, which in turn requires understandable descriptions. \autoref{fig:rq1-understanding} shows that, when asked
    whether they felt like they understood either the analyses (Q10) or their associated benefits (Q21), all participants either agreed or
    strongly agreed, indicating that all of them felt that they understood the scenario.

    When asked to clarify what, if anything, was unclear about the analyses (Q11), eight participants provided responses.
    Three of them were non-answers such as ``N/A'' (R40) or ``-'' (R27). The remaining answers were related to various concerns:
    One participant was unsure about whether they understood everything ``due to it all being new language''~(R20). Indeed, a language
    barrier may exist for some participants because almost all countries we included during the pre-screening stage of our study
    do not have English as their national language. Three participants each referred to an aspect they felt was missing from the analysis
    descriptions provided to them: How long the data will be stored (R8), how exactly the sentiment of messages will be determined for
    \analysisPH, and missing interactivity within the widget of \analysisDE~(although this respondent ``later realized the widget is clickable'' (R5)
    and asked us to ``disregard'' this remark). Finally, one respondent indicated that the ``intensions [sic!] of the manager'' (R11) were unclear. Certainly,
    the manager's true motives may differ from the intended purpose of the analysis. This is a factor that some participants mentioned
    when explaining their initial privacy decisions (see RQ2). Regarding benefits, there were three responses to Q22, all of
    which were non-answers.

    \subsubsection{RQ2 -- How did respondents view each analysis, which factors influenced their initial privacy decisions?}
\begin{figure}[htbp]
      \includegraphics[width=.7\textwidth]{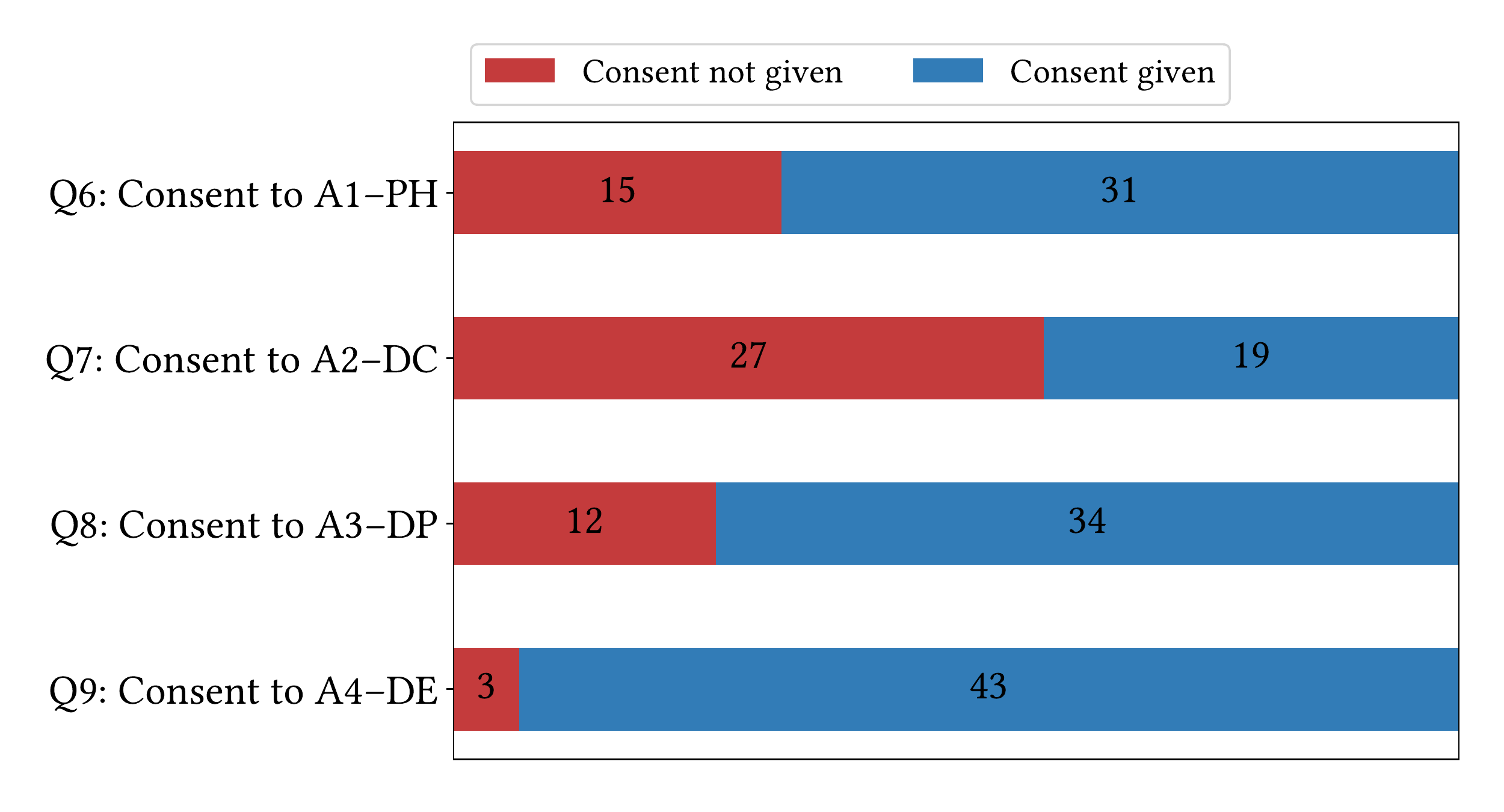}
      \caption{Rates of initial consent to the analyses among our participants.}
      \label{fig:rq2-initial-consent}
      \Description{Depicted are four horizontally-stacked bar graphs representing the consent choices participants initially made in questions Q6 to Q9, distinguishing between ``Consent given'' and ``Consent not given.'' For Q6 (related to \analysisPH{}), 31 respondents gave consent and 15 did not. For Q7 (related to \analysisDC{}), 19 consented, whereas 27 did not. For Q8 (related to \analysisDP{}), 34 consented and 12 did not. Finally, for Q9 (related to \analysisDE{}), 43 consented and 3 did not.}
    \end{figure}

    We first recorded whether each participant would consent to each of the analyses from \refsection{subsec:ex-impl} without having introduced any benefits.
    \autoref{fig:rq2-initial-consent} shows the proportion of participants giving consent per analysis. As can be seen, there are significant differences between
    the analyses: While only three participants stated that they would not consent to \analysisDE{} (developer expertise), 27 participants said that they would
    refuse to provide their data for the purposes of \analysisDC{}  (developer collaboration). With fifteen and twelve non-consenting respondents, the distributions of
    \analysisPH{} (project health) and \analysisDP{}  (developer performance) appear to be similar.

    To understand these differences, we asked participants to explain their choices (Q13 -- Q16). We then assigned codes to each
    answer (see \refsection{subsec:eval-method}) and classified the codes into the following four categories:

    \begin{itemize}
      \item \textbf{Value}: Answers that took into account the value of the analysis for the company or for the data owners themselves.
      This also includes answers that deemed the analysis unnecessary or questioned whether it is suitable to measure what it is
      intended to measure.
      \item \textbf{Sensitivity}: Responses that assessed how sensitive the data involved or how intrusive the analysis is perceived to be.
      \item \textbf{Risks and concerns}: Answers mentioning a perceived risk or the fear of some negative consequence in case
      consent is given.
      \item \textbf{Other}.
    \end{itemize}

    \autoref{tab:rq2-factors} shows how our codes are distributed among the responses to the questions Q13 -- Q16.
    Immediately, we can observe differences between the analyses: \analysisDC{}  (Q14), which our participants were the least willing to
    consent to, was seen as particularly intrusive and sensitive due to the inclusion of private messages in its metric,
    with one participant remarking that ``private messages should remain private''~(R33). To a lesser extent, this also applies to
    \analysisPH{}  (Q13): Multiple participants state that they would not like their messages to be processed for the purposes of
    sentiment analysis as they perceive it to be intrusive. We also observe that \analysisDC{}  had the highest number of answers
    that questioned its validity, typically because they thought the information was not suited to derive the intended insights.
    E.g., ``I don't believe that data from a computer can evaluate the reasons''~(R18) or ``the number of times Developer A mentioned Developer B
    in their messages may not directly correspond to the health of the project''~(R12).

    \begin{table}[htbp]
      \caption{Codes assigned to the answers to Q13 -- Q16 (relating to each analysis).
      Examples can be found in \hyperref[appendix:codes]{Appendix B}.}
      \begin{tabular}{lp{5cm}cccc}
        \toprule
        \textbf{Category}                 & \textbf{Code}                              & \textbf{\analysisPH} & \textbf{\analysisDC} & \textbf{\analysisDP} & \textbf{\analysisDE} \\
        \midrule
        \multirow{5}{*}{Value}            & \makecell[l]{analysis valuable for\\ company / management} & 22           & 11           & 19           & 28           \\
                                          & questions validity of analysis             & 5            & 14           & 9            & 3            \\
                                          & analysis valuable for data owner           & 9            & 6            & 8            & 7            \\
                                          & unnecessary analysis                       & 2            & 4            & 1            & 0            \\
                                          & objective analysis                         & 0            & 1            & 3            & 0            \\
        \midrule
        \multirow{5}{*}{Sensitivity}      & intrusive analysis / sensitive data        & 13           & 18           & 4            & 0            \\
                                          & no intrusive analysis / sensitive data     & 7            & 2            & 3            & 6            \\
                                          & right of the company                       & 1            & 0            & 3            & 6            \\
                                          & data already available                     & 1            & 1            & 1            & 3            \\
                                          & common practice                            & 0            & 0            & 2            & 1            \\
        \midrule
        \multirow{8}{*}{\makecell[l]{Risks and\\concerns}} & no risks                                   & 3            & 2            & 5            & 6            \\
                                          & risk of data misusage                      & 3            & 7            & 5            & 1            \\
                                          & feeling watched                            & 2            & 2            & 4            & 1            \\
                                          & fear of negative consequences              & 0            & 3            & 6            & 0            \\
                                          & risk of misinterpretation                  & 1            & 2            & 3            & 0            \\
                                          & requires consent from all parties          & 2            & 1            & 1            & 1            \\
                                          & monitoring influences behavior             & 1            & 1            & 2            & 0            \\
                                          & unsafe                                     & 1            & 1            & 0            & 1            \\
        \midrule
        \multirow{5}{*}{Other}            & nothing to hide                            & 1            & 1            & 3            & 1            \\
                                          & unsure whether to consent                  & 3            & 2            & 0            & 1            \\
                                          & analysis considered disallowed             & 1            & 0            & 0            & 0            \\
                                          & potential peer pressure                    & 0            & 0            & 1            & 0            \\
                                          & misunderstood consent                      & 0            & 0            & 1            & 0            \\
        \midrule
        \makecell[l]{No code\\ assigned}                 & --                                         & 0            & 0            & 1            & 0            \\
        \bottomrule
        \end{tabular}
      \label{tab:rq2-factors}
    \end{table}

    On the other hand, \analysisDE~(Q16), the analysis that had the highest rate of consent, was not referred to by any participant as very
    sensitive -- instead, some participants argued that it is the right of the company (or that it is common practice for a company) to collect such data if
    management chooses to do so. In addition to this, most respondents highlighted the usefulness of the analysis to the company, e.g.
    through being able to assist employees better or by providing managers with the ability to ``to better plan the work''~(R44).
    Sensitivity-wise, \analysisDP~(Q15) is judged similarly, though fewer participants noted its use to the company compared to \analysisDE.

    Each analysis had a set of perceived risks and concerns associated with it. Multiple participants were concerned
    that their data may be misused by their manager (e.g. to persecute or terminate them), that it may be interpreted wrongly, or otherwise
    lead to negative consequences for employees. As an example, one participant remarks for \analysisDP{} that they would ``feel constantly watched''~(R3)
    and are afraid of being ``punished by the employer'' if their ``performance [...] is slightly worse than the average''~(R3). Indeed, the fear of
    misuse and negative consequences was the most common among \analysisDP~as it intends to measure the performance of employees. This, as another participant remarks,
    could be used to ``justify bad performance reviews.''~(R5) An additional concern mentioned by some respondents is that
    ``employees would just act differently [...] knowing they are being monitored,''~(R14) which may be compounded by the fear of negative
    consequences if they do not have statistics that their manager approves of. For instance, ``employees could do as much as possible
    but with the lowest quality''~(R42) to increase the number of tasks they complete throughout the week, thus improving their statistics
    as presented by the widget of \analysisDP.

    Other noteworthy points raised in answers were a ``nothing to hide'' attitude towards some analyses, e.g.
    ``If you are a productive person,
    you should have no qualms about this information being collected,''~(R41) or a fear of peer pressure regarding consent to analyses because
    ``maybe other colleagues would consent and it would be odd not to also consent''~(R32). Finally, some participants remarked that they
    were unsure about whether to consent (e.g., R17).

    \subsubsection{RQ3 -- Did data owner benefits significantly influence the outcome of privacy decisions in our scenario?}

    During the survey, participants were asked twice about whether they would consent to each analysis in the given scenario,
    first without any benefits, then with our set of data owner benefits. \autoref{tab:rq3-preference-change} shows the overall
    change in consent throughout all participants and across the two variants of the scenario.
    \begin{table}[htbp]
      \caption{Number of participants consenting to the analyses \analysisPH~to \analysisDE{} without (before) and with (after) provided benefits.}
      \begin{tabular}{lccc}
      \toprule
                       & \textbf{After: Consent} & \textbf{After: No consent} & $\Sigma$ \\
                       \midrule\textbf{\analysisPH} & & & \\
      \textbf{Before: Consent}    & 26                      & 5                          & 31       \\
      \textbf{Before: No consent} & 0                       & 15                         & 15       \\
      $\Sigma$                    & 26                      & 20                         & 46       \\
      \midrule
      \textbf{\analysisDC}                 &                         &                            &          \\
      \textbf{Before: Consent}    & 17                      & 2                          & 19       \\
      \textbf{Before: No consent} & 1                       & 26                         & 27       \\
      $\Sigma$                    & 18                      & 28                         & 46       \\
      \midrule
      \textbf{\analysisDP}                 &                         &                            &          \\
      \textbf{Before: Consent}    & 31                      & 3                          & 34       \\
      \textbf{Before: No consent} & 1                       & 11                         & 12       \\
      $\Sigma$                    & 32                      & 14                         & 46       \\
      \midrule
      \textbf{\analysisDE}                 &                         &                            &          \\
      \textbf{Before: Consent}    & 41                      & 2                          & 43       \\
      \textbf{Before: No consent} & 1                       & 2                          & 3        \\
      $\Sigma$                    & 42                      & 4                          & 46       \\
      \bottomrule
      \end{tabular}
      \label{tab:rq3-preference-change}
      \end{table}
    Just from looking at the data,
    we can already see that there is was most probably no significant impact of our introduced benefits on the data sharing consent.
    To confirm our suspicions, we perform McNemar's exact test~\cite{Fay2010} for each analysis and receive
    p-values of 0.0625, 1, 0.625, and 1 for \analysisPH{}, \analysisDC{}, \analysisDP{}, and \analysisDE{} respectively. Out of these, only the p-value for \analysisPH{} approaches statistical
    significance at $\alpha$ = 0.05 -- however, a closer inspection of the reasoning behind these changes reveals that one participant may have misunderstood
    the second scenario, revoking their consent only because their ``boss would be able to see much more than [they] would,''~(R46) which
    would have been the case in the first scenario as well. Accounting for this would increase the p-value for \analysisPH{} to 0.125. In any event,
    our observations do not yield evidence of a statistically significant influence on the outcome of participants' decisions, positive
    or negative. Despite this, the number of non-consenting respondents increases for every analysis, which
    indicates that the introduced benefits could raise concerns among employees. In the following research question, we consider this possibility when analyzing the reasoning for respondents' decision.

    \subsubsection{RQ4 -- How did participants perceive the introduced data owner benefits?}

    Participants were asked to explain why (or why not) the introduced benefits affected their decision-making process (Q24 -- Q27).
    \autoref{tab:rq4-factors} contains the distribution of codes across analyses. As with RQ2, the codes are grouped into
    four categories:

    \begin{itemize}
      \item \textbf{Value}: Answers that assessed the usefulness or the appeal of the introduced benefits.
      \item \textbf{Risks due to benefits}: Responses indicating some concern introduced by the presence of benefits.
      \item \textbf{Impact on decision}: Answers explaining the impact of benefits on the willingness to consent.
      \item \textbf{Other}.
    \end{itemize}

    \begin{table}[htbp]
      \caption{Codes assigned to the answers to Q24 -- Q27 (related to \analysisPH{} to \analysisDE).}
      \begin{tabular}{llcccc}
        \toprule
        \textbf{Category}                   & \textbf{Code}                              & \textbf{\analysisPH} & \textbf{\analysisDC} & \textbf{\analysisDP} & \textbf{\analysisDE} \\
        \midrule
        \multirow{3}{*}{Value}              & benefit appreciated                        & 15          & 8           & 22          & 23          \\
                                            & benefit not appealing                      & 6           & 12          & 3           & 3           \\
                                            & benefit should be part of the tool         & 0           & 0           & 1           & 2           \\
        \midrule
        \multirow{2}{*}{\makecell[l]{Risks due\\ to benefits}} & benefit negatively influences individual   & 4           & 0           & 3           & 0           \\
                                            & benefit increases tracking \& surveillance & 2           & 2           & 1           & 1           \\
        \midrule
        \multirow{6}{*}{\makecell[l]{Impact on\\ decision}} & benefit not necessary for consent          & 8           & 9           & 9           & 15          \\
                                            & benefit not sufficient                     & 4           & 7           & 3           & 1           \\
                                            & benefit changed decision                   & 4           & 3           & 4           & 3           \\
                                            & no benefits could change decision          & 3           & 6           & 2           & 0           \\
                                            & benefit affirmed decision                  & 1           & 0           & 1           & 2           \\
                                            & benefit does not outweigh risk             & 2           & 1           & 0           & 0           \\
        \midrule
        \multirow{4}{*}{Other}              & reciprocity                                & 2           & 2           & 2           & 1           \\
                                            & nothing to hide                            & 1           & 0           & 0           & 1           \\
                                            & unsure about consent                       & 1           & 0           & 0           & 0           \\
                                            & misunderstands benefit                     & 1           & 0           & 0           & 0           \\
        \midrule
        \makecell[l]{No code\\ assigned}                    & --                                         & 3           & 6           & 4           & 4           \\
        \bottomrule
      \end{tabular}
      \label{tab:rq4-factors}
    \end{table}

    The data reveals insights about the role of benefits in the decision-making process: It was common for participants
    to appreciate the benefits in some way, typically because they believed that the provided information would be interesting
    or even valuable: One answer to Q24 states ``It is interesting to see such reports, it can be useful to me to improve my work''~(R15).
    Occasionally, participants went a step further, arguing that the inclusion of the benefit is ``something that should be done''~(R14) or that
    ``if [their] manager can see it, [they] should see it too''~(R39).
    At the same time, not every participant thought the benefits were useful or appealing, e.g. because they believed that
    they could ``manage to know these stats [themselves]''~(R2). Certainly, the provided benefits give data owners an insight on
    statistics that they could infer from their own work.

    While not every participant explicitly talked about the impact that benefits had on their decisions, the findings in
    \autoref{tab:rq4-factors} are generally in line with the values examined for RQ3 (see \autoref{tab:rq3-preference-change}):
    Most commonly, participants stated that the benefits were not necessary for them to consent (because they had already consented)
    or that they were not sufficient to convince them to change their decision, e.g. because the risks of the analysis were perceived to be
    too large to overcome. Indeed, the data depicted in \autoref{tab:rq3-preference-change}
    shows that most participants did not change their decision even after benefits were introduced. Some participants go even further,
    stating that the analysis is sensitive enough such that no benefit could convince them to consent to it, e.g. because of
    the data used in the analysis: ``giving access to my messages is a no for me'' (R24 on Q24).

    In a few cases, we do observe an impact of benefits on the decision, both positive and negative. Participants who were convinced to
    consent by the benefits were typically interested in the data, e.g. ``I would like to know this information''~(R22 on~Q25), whereas
    participants who changed their decision in the opposite manner voiced concerns about the nature of the benefits.
    The benefits of \analysisPH{} were perceived to be problematic due to the fact that it would reveal the aggregated sentiment of the team
    to every data owner who consented. For one participant, this ``makes it a bigger privacy breach,''~(R14)
    another participant feared ``more negative feelings within the group''~(R27) if someone saw that their own sentiment was below the
    group average. Issues with the benefits associated with \analysisDP{} arose from additional stress or being ``observed and controlled''~(R23).

    \subsubsection{RQ5 -- Overall, did participants prefer the introduced people analytics with or without data owner benefits?}
    At the end of the survey, we asked participants whether they preferred the first variant of the scenario (without data owner benefits)
    or the second variant (with data owner benefits). \autoref{tab:rq5-scenario-preference} shows the results, grouped by the role stated
    (manager, employee, or other). In general, managers appeared to be less appreciative of the second variant, with only four out of nine preferring it compared to 77.4\% of employees. Six participants had no preference either way.

    \begin{table}[htbp]
      \caption{Answers to Q28 (i.e., preferring the scenario with or without benefits), grouped by the answers given to Q4 (role in employment).
      While about 77.4\% of employees preferred the variant with data owner benefits, only about 45.5\% of managers did. Overall, about 71.7\% of respondents preferred the second variant.}
      \begin{tabular}{lccc}
        \toprule
        \textbf{Group (Q4)}            & \textbf{Prefer first variant} & \textbf{Prefer second variant} & \textbf{No preference} \\
        \midrule
        Managers                  & 4                             & 5                              & 2                      \\
        Employees                 & 3                             & 24                             & 4                      \\
        Not applicable            & 0                             & 3                              & 0                      \\
        Unsure                    & 0                             & 1                              & 0                      \\
        \midrule
        $\Sigma$                  & 7                             & 33                             & 6                      \\
        \bottomrule
      \end{tabular}
      \label{tab:rq5-scenario-preference}
    \end{table}

    \subsection{Threats to validity}
    \label{subsec:limitations}

    When interpreting the results of our study, it is important to keep potential threats to validity in mind.
    For one, our sample is not representative and its size is limited, meaning that our findings do not generalize.
    Another threat arises from the way we recruited our participants. Users on Prolific receive monetary compensation
    for their successful participation in a study, which may incentivize them to work through questionnaires faster and
    less thoroughly. To mitigate this risk, we included checks in our survey (Q12, Q21, Q23, Q29) for filtering out
    inattentive participants. Another aspect to consider is the design of our study itself: Our questions rely on
    self-reported data and hypothetical scenarios, not on a real-world workplace setting, which may decrease the
    accuracy of our results. Furthermore, the design of our study assumes that the scenario in which data owners gain
    benefits is likelier than the reversed scenario. As a result, all participants saw the questions in the same order;
    there is no control group. It could therefore be sensible to examine the scenario in which the benefits are not introduced,
    but taken away.

    \section{Related work}
    \label{sec:related-work}

    In this section, we examine related literature by covering two aspects. First, we highlight design approaches that tackle issues similar to the ones our principles address and, if applied to PA, would factor in employees as important stakeholders. Then, we explore literature examining the risks and benefits of PA.

    \subsection{Design approaches involving stakeholders}

    Existing design approaches for interactive systems address problems similar to the one presented in this paper, although they are not
    focused solely on PA. Human-centered design~(HCD)~\cite{ISOHCD}, for instance,
    aims to improve the user experience by actively involving users throughout
    the design and development process. The goal is an overall
    better usability, e.g. through reduced stress for the users of a system~\cite{ISOHCD}.
    However, there are also limitations to significant user involvement. Since users are generally not experienced designers
    themselves, their requests may lead to convoluted designs if followed without careful consideration~\cite{Norman2005}. A related approach
    is participatory design~(PD), which encompasses various techniques~\cite{Muller1993} that
    translate as participatory practices into the software development process~\cite{Muller1997}. PD involves the users of the system as
    important stakeholders during the design process.

    If applied to PA, such approaches would incorporate the interests of data owners and data consumers as stakeholders into the design of the respective PA tool. As our work is intended to be a first step towards improving PA design and the involvement of users does not always lead to better results~\cite{Norman2005}, we have decided against involving stakeholders in the design of our implementation. However, it may be sensible to pursue a combination of either PD or HCD with our approach as a follow-up based on the findings described in \refsection{subsec:results}.

    The interaction of employee participation and different aspects of an organization have been studied from varying angles. For example, \citeauthor{Zhou2019}~\cite{Zhou2019} find that, in the context of human resource management, employee participation can boost organizational innovation~\cite{Zhou2019}. Another example is the work of \citeauthor{LeichtDeobald2019}~\cite{LeichtDeobald2019}, who explore the impact HRM tools akin to PA have on the personal integrity of employees. One of their recommendations is to apply PD. By making managers and employees equals in the design process, all of them will have the opportunity to provide their unique perspectives. This, in turn, improves transparency and may help employees retain their personal integrity when HRM tools are used~\cite{LeichtDeobald2019}.

    \subsection{Investigating the risks and benefits of PA}

    Few works consider the problem of data owner appeal in PA~\cite{Zieglmeier2022}.
    \citeauthor{Zieglmeier2022}~\cite{Zieglmeier2022} approach the topic in a generic fashion and provide a taxonomy
    of appeal strategies covering values, incentives, and benefits. They evaluate their taxonomy theoretically
    in expert interviews. However, their appeal strategies have not yet been practically applied or
    evaluated with employees themselves. We address this research gap by concretizing their concept and implementing the benefit strategy~\cite[Fig. 1]{Zieglmeier2022}.

    \citeauthor{Margherita2022}~\cite{Margherita2022}, in comparison, adopt an organizational view on employee value of PA.
    Based on a systematic literature review, they identify key research topics categorized as enablers, applications, and value of PA, which includes employee value.
    Due to their focus, they naturally present a more theoretical overview, but their work can be a useful foundation.
    For example, they identify that PA can support employee reskilling or provide them with wage transparency~\cite{Margherita2022}, which can be classified as PA benefits.
    Their work does not consider the design of PA, though, and therefore can be seen as complementary.
    \citeauthor{Chatterjee2022}~\cite{Chatterjee2022} also study the risks and benefits of PA.
    Their research is more similar to our work, in that they examine the attitudes of individuals towards PA, taking into account their potential upsides and
    privacy implications. However, their participants were asked to rate several statements regarding PA on a Likert scale~\cite{Chatterjee2022},
    whereas we asked participants to provide their reasoning in free text answers. More importantly, our work addresses the design
    of PA, which is not the primary focus in~\cite{Chatterjee2022}.

    On the other side of the spectrum, \citeauthor{Khan2016}~\cite{Khan2016} and \citeauthor{Giermindl2022}~\cite{Giermindl2022} focus more on the risks and concerns regarding PA.
    Privacy calculus theory suggests that both risks and benefits are relevant for an employee's disclosure decision~\cite{bhave2020privacy, Zieglmeier2022}.
    We should recognize, though, that some risks and ethical issues of PA are arguably too large to be tolerated~\cite[see, e.g.,][]{LeichtDeobald2019}.
    For example, \citeauthor{Khan2016}~\cite{Khan2016} warn that the boundaries of employee monitoring need to be clearly delineated.
    They suggest that the inclusion of, e.g., wearable data or social media monitoring should be critically examined~\cite{Khan2016}.
    The risk of such invasive monitoring is also highlighted by \citeauthor{Giermindl2022}, who note that, in the workplace, the forced adoption of technologies is combined with more severe consequences for an individual.
    Therefore, they warn that the logic from other analytics should not just be applied to PA, but adapted according to those factors~\cite{Giermindl2022}.

    \section{Discussion}

    In this section, we discuss our results in the greater context of people analytics and examine the privacy concerns mentioned by participants. Additionally, we explain what the results could mean for
    the role of data owner benefits in the design of PA and discuss the possibility of involving stakeholders in the design process.

    \subsection{Perceived risks and concerns regarding our analyses}

    The overarching goal of this work is to increase the willingness of data owners to disclose their data. We therefore analyze
    the factors that influenced the disclosure decisions of our respondents.
    While we do find that participants commonly recognize the value of the analyses (for the company or otherwise), there are also
    concerns associated with the analyses. These concerns may (and in our sample, seem to)
    negatively influence the willingness to provide consent in some participants and should be addressed.
    In the following, we summarize the concerns mentioned by participants and explain their implications for the design, implementation,
    and use of PA.

    One concern that respondents took into account is the value an analysis can provide, either to the company or its employees.
    In particular, several participants questioned the validity of analyses itself, either due to its desired insight or how it processes data. For instance, multiple participants disapproved of the use of sentiment analysis in the metric of \analysisPH~(project health). Indeed, as analyses become more sophisticated with the use of artificial intelligence, their underlying reasoning may become less
    transparent~\cite{Giermindl2022}, thus potentially decreasing trust.

    Another concern is the sensitivity of the data involved. This aligns with our privacy model~\cite{Smith2011}: the riskier or more costly a data usage is perceived to be, the less likely someone is to disclose their data. \analysisDC{} (developer collaboration),
    which was considered to be particularly intrusive and questionable due to its use of private messages in its metric,
    had a low consent rate compared to \analysisDE{} (developer expertise), which was generally not seen as intrusive.

    Third, respondents remarked upon the possibility of misuse and misinterpretation by the manager. For instance, the analysis \analysisDP{} (developer performance) could, according to some participants, be used to persecute or
    punish them in some fashion. Of course, the \textit{true} intentions of the manager analyzing the data cannot be captured by
    the description of the purpose of the analysis such as the ones we provided to participants
    (see \hyperref[appendix:analysis-descriptions]{Appendix C}). To deter such misusages, it may be sensible to make all data usages available to the corresponding data owners~\cite[see,~e.g.,][]{Zieglmeier2022a}.

    Finally, respondents also mentioned other negative consequences, commonly related to feeling
    watched or pressured by the analyses, potentially resulting in negative consequences for the well-being of employees.
    The source of this pressure could be peer pressure, as the manager knows who consented to each
    analysis, which could result in non-consenting employees appearing as though they are trying to hide information about their performance.
    Alternatively, the pressure could stem from the analysis itself, e.g. in the case of \analysisDP, employees may feel the need to adapt
    their behavior to the metric of the analysis rather than improve the way they work.

    Ultimately, the concerns mentioned by our respondents relate back to the conceptual limitations discussed in \refsection{sec:approach}.
    Participants remark upon the collected data, the implementation used to process it, and how the manager may interpret the results. This
    highlights the importance of these issues and shows why designers should be transparent about them, for data owners may
    consider them relevant in their decision-making.

    \subsection{The role of data owner benefits}

    Our results suggest that the introduced benefits were not sufficient to significantly
    influence the decisions of our participants. This could mean that the specific benefits we
    defined in \refsection{subsec:ex-impl} are ill-suited and a different set of benefits could have had a more significant impact --
    after all, only 11 of 184 answers to the questions Q24~--~Q27 state that no benefit would have been able to change the decision.
    This indicates that not the concept of benefits in general, but their specific instantiation in our case were a factor in their
    effectiveness. Moreover, we find that several respondents whose decision was unaffected still appreciated having data owner benefits, with over 70\% of
    participants preferring the variant of the scenario that includes them. Therefore, it is likely that data owner
    benefits have a place within PA regardless of their power to affect disclosure decisions, if only to increase the acceptance
    of PA among employees. Even so, PA developers should be careful not to
    conceptualize benefits that increase privacy concerns (e.g. some participants took issue with the fact that the benefit of
    \analysisPH~would provide aggregated data about them to their colleagues).

    On a related note, managers were more skeptical of the variant with data owner benefits, with less than half of them preferring it
    over the base scenario. While the remaining data does not yield any conclusions about a unique managerial perspective, this could
    be one possible aspect for future studies. As managers may inhabit the role of a data owner as well as a data consumer, it may be interesting to differentiate the managerial perspective more based on this aspect.

    \subsection{Addressing conflicting interests by involving stakeholders in the design}

    Our results indicate that, contrary to our expectations, the benefits did not sufficiently alter the disclosure
    decisions of data owners. Furthermore, there were several concerns associated with the analyses themselves.
    One potential issue could be the conflicting interests of the different stakeholders.
    According to \citeauthor{Tursunbayeva2022}~\cite{Tursunbayeva2022}, managers use PA to achieve strategic goals, whereas employees may have concerns regarding how their data is analyzed and used.
    This contrast creates two distinct perspectives that must both be considered, which is why \citeauthor{Tursunbayeva2022} recommend including employees in PA projects to prevent unethical PA practices.
    \citeauthor{Khan2016}~\cite{Khan2016} also acknowledge a divide in how different stakeholders view PA.
    In particular, they state that the workforce may respond negatively to PA should their concerns remain unaddressed.
    These issues may be one potential explanation for our findings.

    To address and resolve employee concerns at an early stage in development, the design approach of PD recommends discussing and resolving conflicting interests early in the design process.
    This should be done by involving all stakeholders, in our case both data owners and data consumers, in the development of PA tools~\cite{Muller1993, robertson2013participatory}.
    Combining PD with our suggested approach may lead
    to a design that further increases the appeal to data owners.
    Additionally and arguably more importantly, as data owners would be involved at an early stage, their concerns could be addressed in the design.
    This allows developers to verify early whether their implementation meets our design principles, based on stakeholder feedback.

    \section{Conclusion}

    Our contribution in this paper is threefold. First, we explore the concept of
    benefits as an appeal strategy and, based on recommendations from existing literature, devise a set of principles for the
    data owner benefit-driven design of people analytics. Then, we illustrate how our principles could be put into practice
    by providing a set of analyses with corresponding data owner benefits and describing an exemplary implementation.
    Finally, we evaluate our principles by analyzing and discussing the results of the user study we have conducted. We examine
    the attitudes of our participants towards the provided analyses and benefits and assess the impact of
    data owner benefits on the privacy decisions in the fictitious scenario of our study.

    While our findings do not provide evidence that indicates a significant impact on privacy decisions, we consider them valuable
    as they reveal various potential privacy concerns that employees may encounter in the presence of PA. Our results further
    show that data owner benefits were generally appreciated among the respondents of our survey. We believe that data owner benefits are
    likely a sensible addition to the design of PA, albeit not one that is capable of addressing every perceived risk.
    We hope our study inspires further research into both privacy concerns in PA and the use of data owner benefits
    to improve the appeal of PA for those subjected to their analyses.

    \bibliographystyle{ACM-Reference-Format}
    \bibliography{literature.bib}

    \clearpage
    \appendix

    \section{Questionnaire}
    \label{appendix:questionnaire}

    Our questionnaire contained the following pages and questions:

    \subsection{Demographics}

    \begin{itemize}
      \item \textbf{Q1} (single choice): How old are you?
      \begin{itemize}
        \item 24 or younger
        \item 25 -- 34
        \item 35 -- 44
        \item 45 -- 54
        \item 55 -- 64
        \item 65 or older
      \end{itemize}
      \item \textbf{Q2} (single choice): What is your gender?
      \begin{itemize}
        \item Male
        \item Female
        \item Other
        \item Prefer not to say
      \end{itemize}
      \item \textbf{Q3} (single choice): Which of these best describes your current employment status?
      \begin{itemize}
        \item Unemployed
        \item Student
        \item Employed
        \item Self-employed
        \item Retired
        \item Other
      \end{itemize}
      \item \textbf{Q4} (single choice): If applicable, would you consider your role to be closer to that of a manager or that of an employee?
      \begin{itemize}
        \item Manager
        \item Employee
        \item Unsure
        \item Not applicable
      \end{itemize}
      \item \textbf{Q5} (single choice): If applicable, would you describe your work as closer to a white-collar or a blue-collar job?
      \begin{itemize}
        \item White-collar (e.g. a desk job, administrative work)
        \item Blue-collar (requiring manual labor)
        \item Unsure
        \item Not applicable
      \end{itemize}
    \end{itemize}

    \subsection{Base scenario description}

\begin{figure}[htbp]
      \fbox{\includegraphics[width=.7\textwidth]{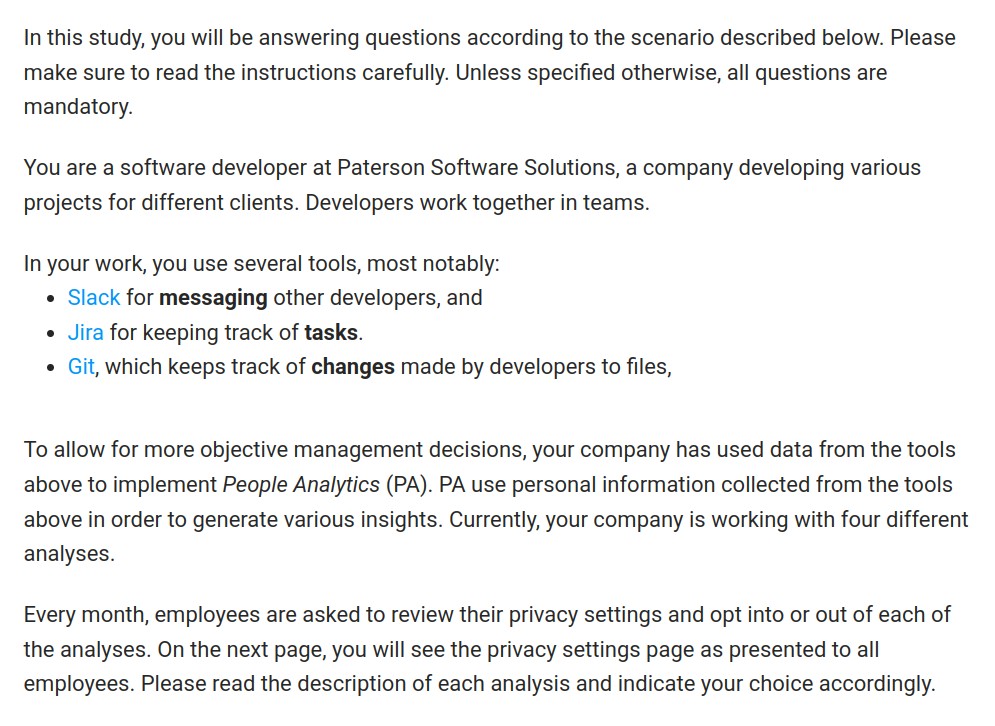}}
      \caption{The scenario description presented to participants of our survey.}
      \label{fig:base-scenario-description}
      \Description{The description reads, ``In this study, you will be answering questions according to the scenario described below. Please make sure to read the instructions carefully. Unless specified otherwise, all questions are mandatory.
      You are a software developer at Paterson Software Solutions, a company developing various projects for different clients. Developers work together in teams. In your work, you use several tools, most notably: Slack for messaging other developers, and Jira for keeping track of tasks. Git, which keeps track of changes made by developers to files. To allow for more objective management decisions, your company has used data from the tools above to implement People Analytics (PA). PA use personal information collected from the tools above in order to generate various insights. Currently, your company is working with four different analyses.
      Every month, employees are asked to review their privacy settings and opt into or out of each of the analyses. On the next page, you will see the privacy settings page as presented to all employees. Please read the description of each analysis and indicate your choice accordingly.''}
    \end{figure}

    Participants were then shown the scenario description depicted in \autoref{fig:base-scenario-description}.

    \subsection{Consent in the base scenario}

    For each analysis, the participants were presented with a description (see \hyperref[appendix:analysis-descriptions]{Appendix C}) asked to indicate their preference:

    \begin{itemize}
      \item Please read the description of each analysis and indicate if you would consent to this use of your data.
      \begin{itemize}
        \item \textbf{Q6} (yes / no, referring to A1): I would consent to the use of my data for the purposes of this analysis in the described fictitious scenario.
        \item \textbf{Q7} (yes / no, referring to A2): I would consent to the use of my data for the purposes of this analysis in the described fictitious scenario.
        \item \textbf{Q8} (yes / no, referring to A3): I would consent to the use of my data for the purposes of this analysis in the described fictitious scenario.
        \item \textbf{Q9} (yes / no, referring to A4): I would consent to the use of my data for the purposes of this analysis in the described fictitious scenario.
      \end{itemize}
    \end{itemize}

    \subsection{Questions about the base scenario}

    \begin{itemize}
      \item \textbf{Q10} (Likert scale): How much do you agree with the statement 'I felt like I understood all of the analyses (A1 to A4)'?
      \begin{itemize}
        \item Strongly agree
        \item Agree
        \item Neither agree nor disagree
        \item Disagree
        \item Strongly disagree
      \end{itemize}
      \item \textbf{Q11} (free text, optional): If anything, what was unclear?
      \item \textbf{Q12} (attention check, single choice): For this question, please choose the option 'No, I definitely do not think so'- it is a test to ensure that you are paying attention.
      \begin{itemize}
        \item Yes, that would definitely be helpful
        \item Yes, but I would not use it
        \item No, I definitely do not think so
        \item I am unsure
      \end{itemize}
    \end{itemize}

    \begin{itemize}
      \item Briefly explain your privacy choices - why did you (not) consent to each analysis? (Each answer should be approximately one to three sentences.)
      \begin{itemize}
        \item \textbf{Q13} (free text): A1 - Project health
        \item \textbf{Q14} (free text): A2 - Developer collaboration
        \item \textbf{Q15} (free text): A3 - Developer performance
        \item \textbf{Q16} (free text): A4 - Developer expertise
      \end{itemize}
      \item \textbf{Note}: To allow participants to reflect upon their choice, their privacy choice was appended to each question (\textbf{Q13 -- Q16}), e.g. A1 - Project health \textit{(consent given)} if the respondent had indicated that they would consent to \analysisPH{}  in question \textbf{Q6}.
    \end{itemize}

    \subsection{Scenario change description}

\begin{figure}[htbp]
      \fbox{\includegraphics[width=.7\textwidth]{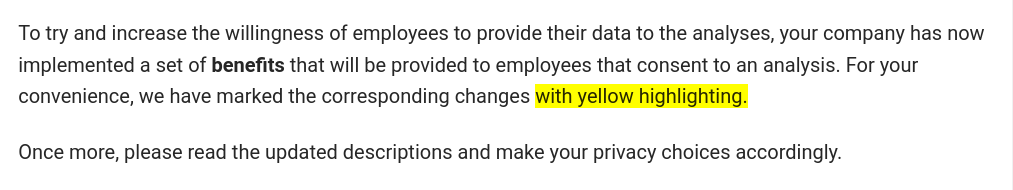}}
      \caption{Description of the change made to the scenario.}
      \label{fig:changed-scenario-description}
      \Description{The description reads, ``To try and increase the willingness of employees to provide their data to the analyses, your company has now implemented a set of benefits that will be provided to employees that consent to an analysis. For your convenience, we have marked the corresponding changes with yellow highlighting.
      Once more, please read the updated descriptions and make your privacy choices accordingly.''}
    \end{figure}

    The scenario was now changed to include data owner benefits -- the concrete description is shown in \autoref{fig:changed-scenario-description}.

    \subsection{Consent in the changed scenario}

    This time, respondents were shown updated variants of the analysis descriptions (\hyperref[appendix:analysis-descriptions]{Appendix C}) and asked:

    \begin{itemize}
      \item Once again, please indicate whether you would consent to each analysis. Changes are highlighted in yellow.
      \begin{itemize}
        \item \textbf{Q17} (yes / no, referring to A1): I would consent to the use of my data for the purposes of this analysis in the described fictitious scenario.
        \item \textbf{Q18} (yes / no, referring to A2): I would consent to the use of my data for the purposes of this analysis in the described fictitious scenario.
        \item \textbf{Q19} (yes / no, referring to A3): I would consent to the use of my data for the purposes of this analysis in the described fictitious scenario.
        \item \textbf{Q20} (yes / no, referring to A4): I would consent to the use of my data for the purposes of this analysis in the described fictitious scenario.
      \end{itemize}
    \end{itemize}

    \subsection{Questions about the changed scenario}

    \begin{itemize}
      \item \textbf{Q21} (Likert scale): How much do you agree with the statement 'I felt like I understood what benefits I would receive if I consented to an analysis'?
      \begin{itemize}
        \item Strongly agree
        \item Agree
        \item Neither agree nor disagree
        \item Disagree
        \item Strongly disagree
      \end{itemize}
      \item \textbf{Q22} (free text, optional): If not, what was unclear?
      \item \textbf{Q23} (Likert scale): Given the statement 'I understood what benefits I would receive if I consented to an analysis', how much do you agree?
      \begin{itemize}
        \item Strongly agree
        \item Agree
        \item Neither agree nor disagree
        \item Disagree
        \item Strongly disagree
      \end{itemize}
    \end{itemize}

    \begin{itemize}
      \item Once more, briefly explain your privacy choices. Did the introduced benefits affect your decision? If so, how? If not, why not? (Each answer should be approximately one to three sentences.)
      \begin{itemize}
        \item \textbf{Q24} (free text): A1 - Project health
        \item \textbf{Q25} (free text): A2 - Developer collaboration
        \item \textbf{Q26} (free text): A3 - Developer performance
        \item \textbf{Q27} (free text): A4 - Developer expertise
      \end{itemize}
      \item \textbf{Note}: Once more, participants were presented with their previous choices, e.g. A1 - Project health \textit{(consent not given $\rightarrow$ consent given)} if they had indicated consent in \textbf{Q17} but not in \textbf{Q6}.
    \end{itemize}

    \subsection{Preference and final remarks}

    \begin{itemize}
      \item \textbf{Q28} (single choice): All in all, which variant of the scenario would you prefer?
      \begin{itemize}
        \item First variant (without employee benefits)
        \item Second variant (with employee benefits)
        \item No preference
      \end{itemize}
      \item \textbf{Q29} (attention check, Likert scale without neutral option): How much do you agree with the statement 'I breathe at least once per day.'?
      \begin{itemize}
        \item Strongly agree
        \item Agree
        \item Disagree
        \item Strongly disagree
      \end{itemize}
      \item \textbf{Q30} (free text, optional): Are there any other remarks you would like to make?
    \end{itemize}

    \section{Qualitative analysis: codes}
    \label{appendix:codes}

    \autoref{tab:appendix-codes-analyses} and \autoref{tab:appendix-codes-benefits} contain the codes assigned to answers
    during our qualitative analysis.

    \begin{longtable}{p{3.5cm}p{5cm}cc}
      \caption{Codes assigned to the answers to questions Q13 -- Q16.} \label{tab:appendix-codes-analyses}\\
      \toprule
      \textbf{Code}                              & \textbf{Example excerpt}                                                                                                   & \textbf{Participant} & \textbf{Analysis} \\
      \endfirsthead
      \multicolumn{4}{c}{\small\tablename\ \thetable{}~(continued). Codes assigned to the answers to questions Q13 -- Q16.} \\
      \toprule
      \textbf{Code}                              & \textbf{Example excerpt}                                                                                                   & \textbf{Participant} & \textbf{Analysis} \\
      \midrule
      \endhead
      \bottomrule
      \multicolumn{4}{c}{\textbf{-- Continued on the following page --}}
      \endfoot
      \bottomrule
      \endlastfoot
      \midrule
      analysis valuable for company / management & ``I do find this to be an invaluable tool for the manager.''                                                               & R5                   & \analysisDP               \\
      questions validity of analysis             & ``I think the number of private messages is not authoritative''                                                            & R44                  & \analysisDC                \\
      analysis valuable for data owner           & ``This helps me get better training offers and better assignments''                                                        & R11                  & \analysisDE{}                \\
      unnecessary analysis                       & ``I believe that this collection of such data is completely unnecessary''                                                  & R38                  & \analysisDC                \\
      objective analysis                         & ``This is the most objective way to measure my work.''                                                                     & R17                  & \analysisDP               \\
      \midrule
      intrusive analysis / sensitive data        & ``I do not want my (or anyone else's) messages to be analyzed in any way.''                                                & R12                  & \analysisPH                \\
      no intrusive analysis / sensitive data     & ``it does not seem intrusive on sensitive/personal matters for the workers''                                               & R5                   & \analysisPH                \\
      right of the company                       & ``It is not unreasonable to collect such data if the employer wishes.''                                                    & R41                  & \analysisPH                \\
      data already available                     & ``The number of meetings and Slack messages made between colleagues is easily accessible within a company anyway.''        & R32                  & \analysisPH                \\
      common practice                            & ``It is a standard procedure at many workplaces to measure workers' performance [...]''                                  & R12                  & \analysisDP               \\
      \midrule
      no risks                                   & ``I was happy to consent for no other reason that it would have no adverse affects.''                                      & R41                  & \analysisPH                \\
      risk of data misusage                      & ``I believe that this collection of such data [...] may be misappropriated.''                                            & R38                  & \analysisDC                \\
      feeling watched                            & ``I'd feel constantly watched.''                                                                                           & R3                   & \analysisDP               \\
      fear of negative consequences              & ``it could be detrimental to the mental well-being of employees''                                                          & R37                  & \analysisDC                \\
      risk of misinterpretation                  & ``The company could draw incorrect conclusions from this data''                                                            & R37                  & \analysisPH                \\
      requires consent from all parties          & ``I did not include this analysis as some employees might not wish to do so''                                              & R37                  & \analysisPH                \\
      monitoring influences behavior             & ``it feels like employees would just act differently in the text chain knowing they are being monitored''                  & R14                  & \analysisPH                \\
      unsafe                                     & ``I wouldn't feel safe allowing this.''                                                                                    & R23                  & \analysisDE{}                \\
      \midrule
      nothing to hide                            & ``If you are a productive person, you should have no qualms about this information being collected.''                     & R41                  & \analysisDP               \\
      unsure about consent                       & ``I was unsure to give consent.''                                                                                         & R17                  & \analysisPH                \\
      analysis considered disallowed             & ``It breaks privacy''                                                                                                      & R19                  & \analysisPH                \\
      potential peer pressure                    & ``maybe other colleagues would consent and it would be odd not to also consent.''                                          & R32                  & \analysisDP               \\
      misunderstood consent                      & ``The productivity of one member/employee can't be measured just by the number of tasks in a week, [...]'' -- yet gives consent & R31                  & \analysisDP              \\
    \end{longtable}

    \begin{table}[htbp]
      \caption{Codes assigned to the answers to questions Q24 -- Q27.}
      \begin{tabular}{p{3.5cm}p{5.5cm}cc}
        \toprule
        \textbf{Code}                                       & \textbf{Example excerpt}                                                                                                        & \textbf{Participant} & \textbf{Analysis} \\
        \midrule
        benefit appreciated                        & ``I think it is interesting to see such reports, it can be useful to me''                                      & R15         & \analysisPH       \\
        benefit not appealing                      & ``the information I would be provided is pretty useless''                                                      & R12         & \analysisPH       \\
        benefit should be part of the tool         & ``This is a good benefict [sic!] and something that should be done.''                                      & R14         & \analysisDP      \\
        \midrule
        benefit negatively influences individual   & ``Having this benefit would stress me out even more. ''                                                        & R3          & \analysisDP      \\
        benefit increases tracking \& surveillance & ``If anything, having the data [...] available to everyone who consents makes it a bigger privacy breach.'' & R14         & \analysisPH       \\
        \midrule
        benefit not necessary for consent          & ``I will share the information with or without the benefit''                                                   & R9          & \analysisDE{}       \\
        benefit not sufficient                     & ``The benefits are not enough.''                                                                               & R40         & \analysisDP      \\
        benefit changed decision                   & ``i have changed my idea just for the plus tool that would be curious to se [sic!]''                       & R21         & \analysisDP      \\
        no benefits could change decision          & ``I dont think any benefits would have convinced me to accept these terms.''                                   & R7          & \analysisPH       \\
        benefit affirmed decision                  & ``that extra tool for me has confirmed my choice''                                                             & R21         & \analysisPH       \\
        \makecell[l]{benefit does not\\outweigh risk}             & ``they do not change my opinion [...], I still think it is very intrusive''                                & R13         & \analysisPH       \\
        \midrule
        reciprocity                                & ``If my manager can see it, I should see it too''                                                              & R39         & \analysisPH       \\
        nothing to hide                            & ``I dont really care what they will do''                                                                      & R1          & \analysisPH       \\
        unsure about consent                       & ``I'm still skeptical about the usefullness [sic!] of chat monitoring - but still agreed''                 & R25         & \analysisPH       \\
        misunderstands benefit                     & ``I did not like the fact that I would not receive some of the project benefits...''                             & R40         & \analysisPH      \\
        \bottomrule
        \end{tabular}
      \label{tab:appendix-codes-benefits}
    \end{table}

    \clearpage
    \section{Analysis descriptions}
    \label{appendix:analysis-descriptions}

    The images at the end of this document depict the analysis descriptions as they were provided to the participants of our questionnaire. Each description has two variants -- one without any data owner benefits, one with data owner benefits.

    \begin{figure}[H]
      \fbox{\includegraphics[width=.95\textwidth]{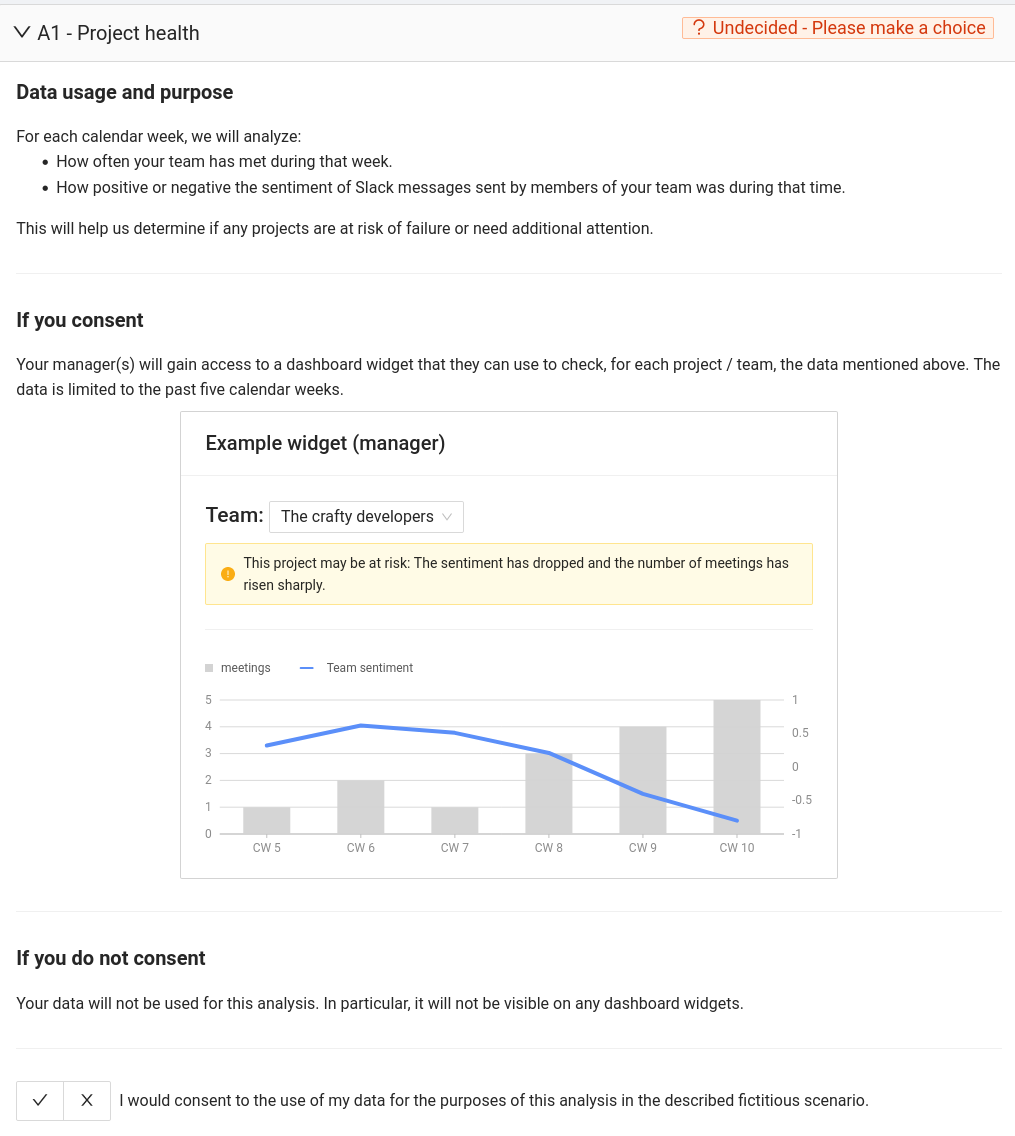}}
      \caption{Description of \analysisPH{} (project health) without data owner benefits in our exemplary implementation.}
      \Description{A screenshot of a website. At the top, a list entitled ``data usage and purpose,'' noting that the analysis will encompass how often the team met and how the sentiment of their Slack messages is. Below, a section ``If you consent,'' describing the analysis textually and presenting a screenshot. It is the analysis shown in Figure 4. Finally, a section ``If you do not consent,'' describing that data won't be used then. At the bottom, two buttons, one to confirm, one to reject sharing.}
      \label{fig:description-a1-base}
    \end{figure}

    \begin{figure}[H]
      \fbox{\includegraphics[width=.95\textwidth]{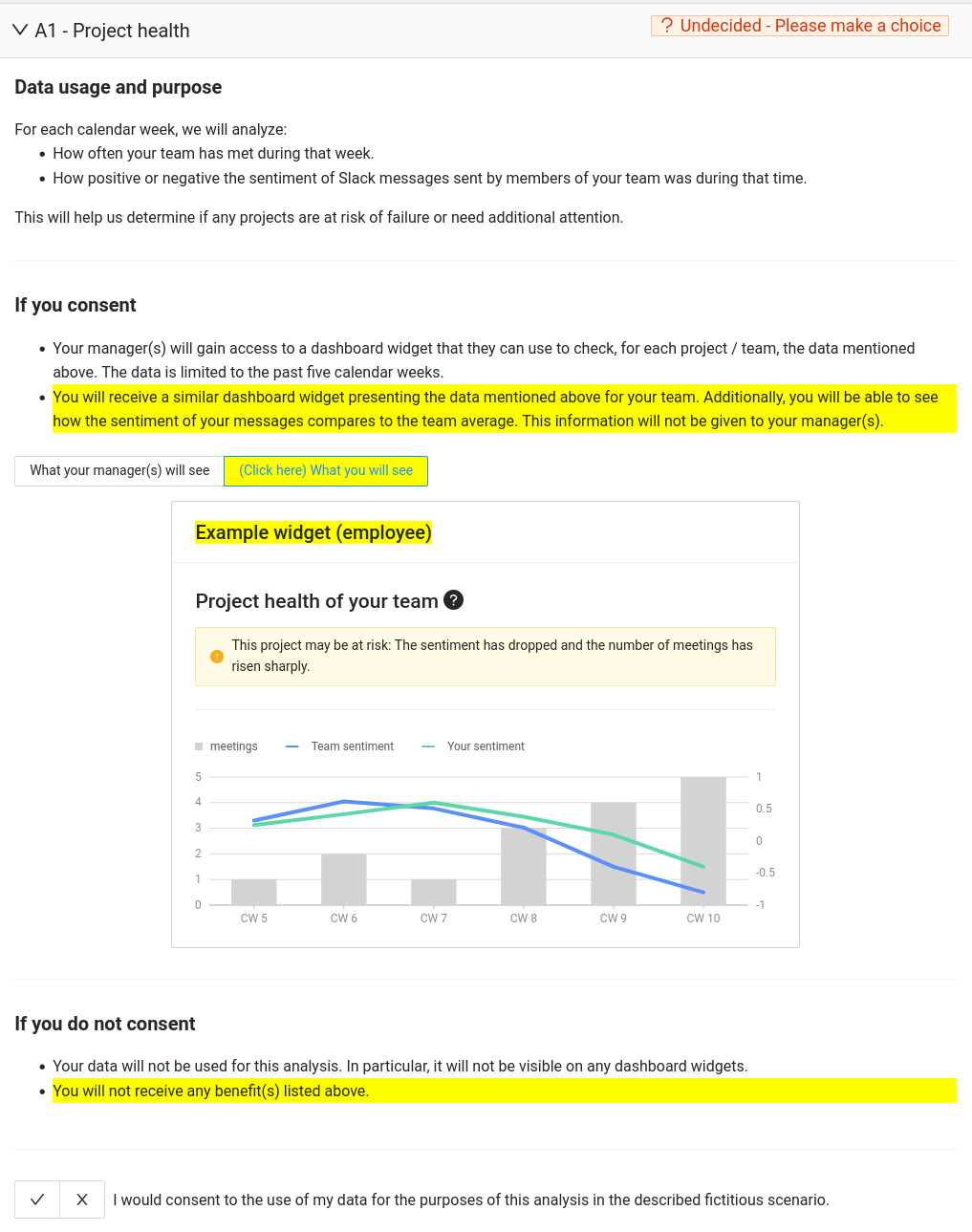}}
      \caption{Description of \analysisPH{} (project health) with data owner benefits in our exemplary implementation.}
      \Description{A screenshot of a website. It is the same as in Figure 12, with the only change being that now, the section ``If you consent'' additionally lists benefits, namely that the user will also receive insights about themselves. Accordingly, the widget also shows a second graph for their own sentiment.}
      \label{fig:description-a1-changed}
    \end{figure}

    \begin{figure}[H]
      \fbox{\includegraphics[width=.95\textwidth]{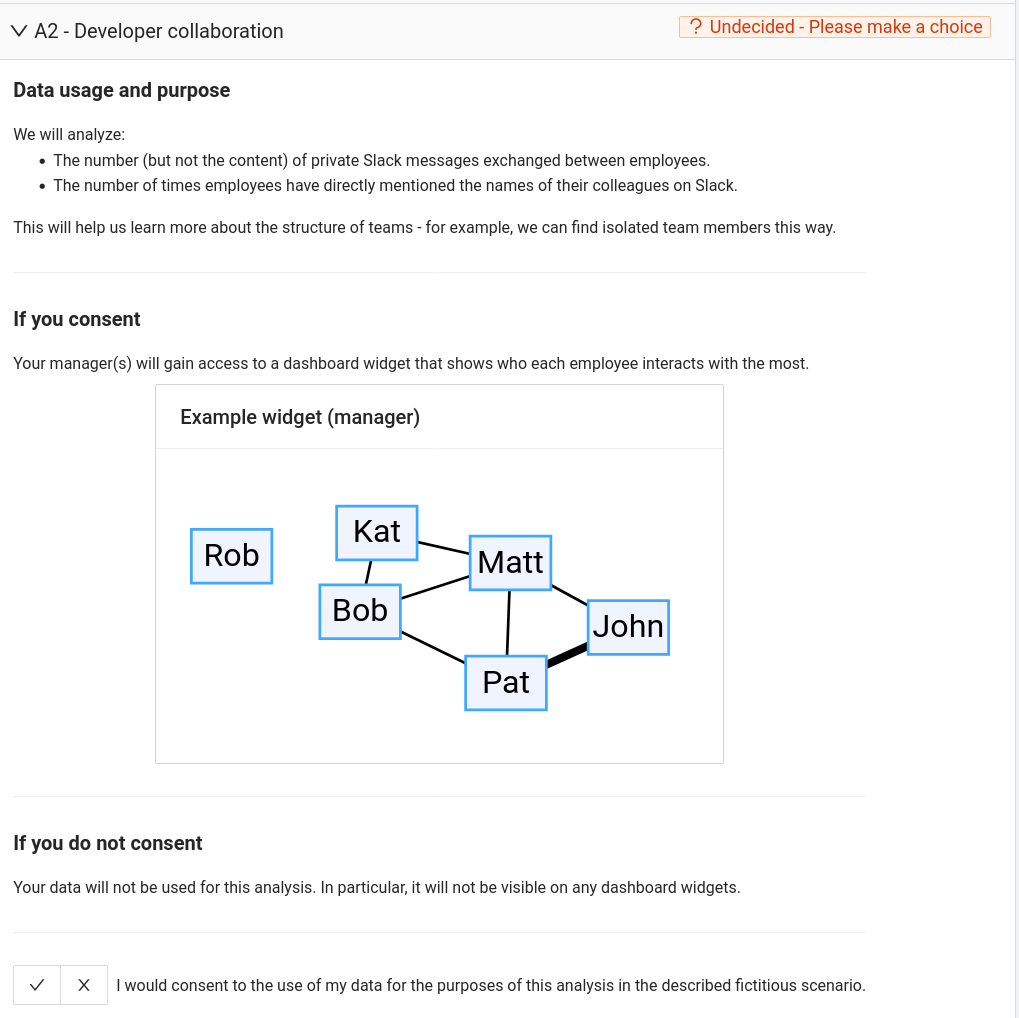}}
      \caption{Description of \analysisDC{} (developer collaboration) without data owner benefits in our exemplary implementation.}
      \Description{A screenshot of a website. At the top, a list entitled ``data usage and purpose,'' noting that the number, not the content, of private Slack messages between employees and the number of their ``mentions'' of other employees will be analyzed. Below, a section ``If you consent,'' describing the analysis textually and presenting a screenshot. It is the analysis shown in Figure 5. Finally, a section ``If you do not consent,'' describing that data won't be used then. At the bottom, two buttons, one to confirm, one to reject sharing.}
      \label{fig:description-a2-base}
    \end{figure}

    \begin{figure}[H]
      \fbox{\includegraphics[width=.95\textwidth]{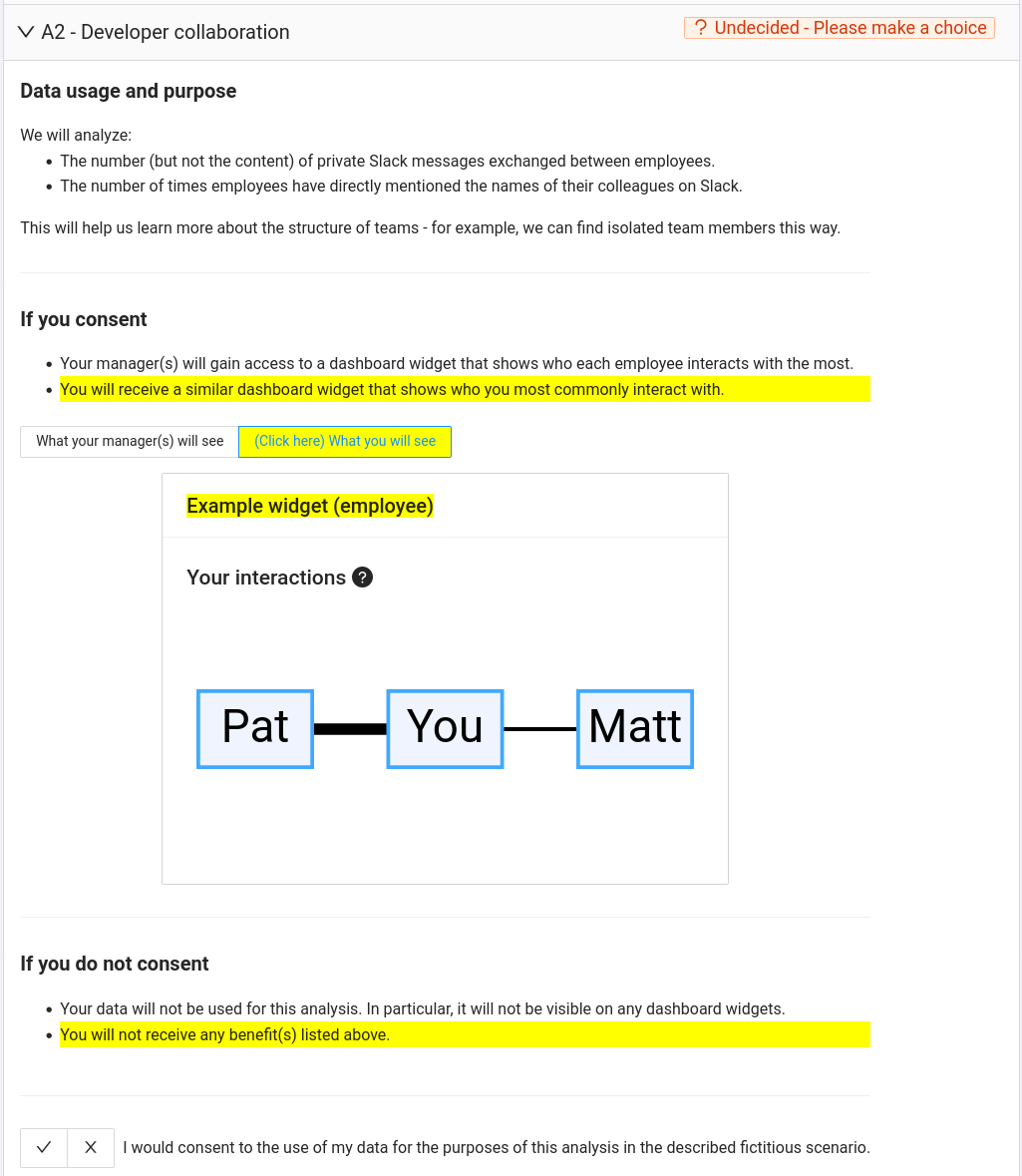}}
      \caption{Description of \analysisDC{} (developer collaboration) with data owner benefits in our exemplary implementation.}
      \Description{A screenshot of a website. It is the same as in Figure 14, with the only change being that now, the section ``If you consent'' additionally lists benefits, namely that the user will be able to see a graph of who they interact with. Accordingly, the widget shows a network graph with their own communication partners.}
      \label{fig:description-a2-changed}
    \end{figure}

    \begin{figure}[H]
      \fbox{\includegraphics[width=.95\textwidth]{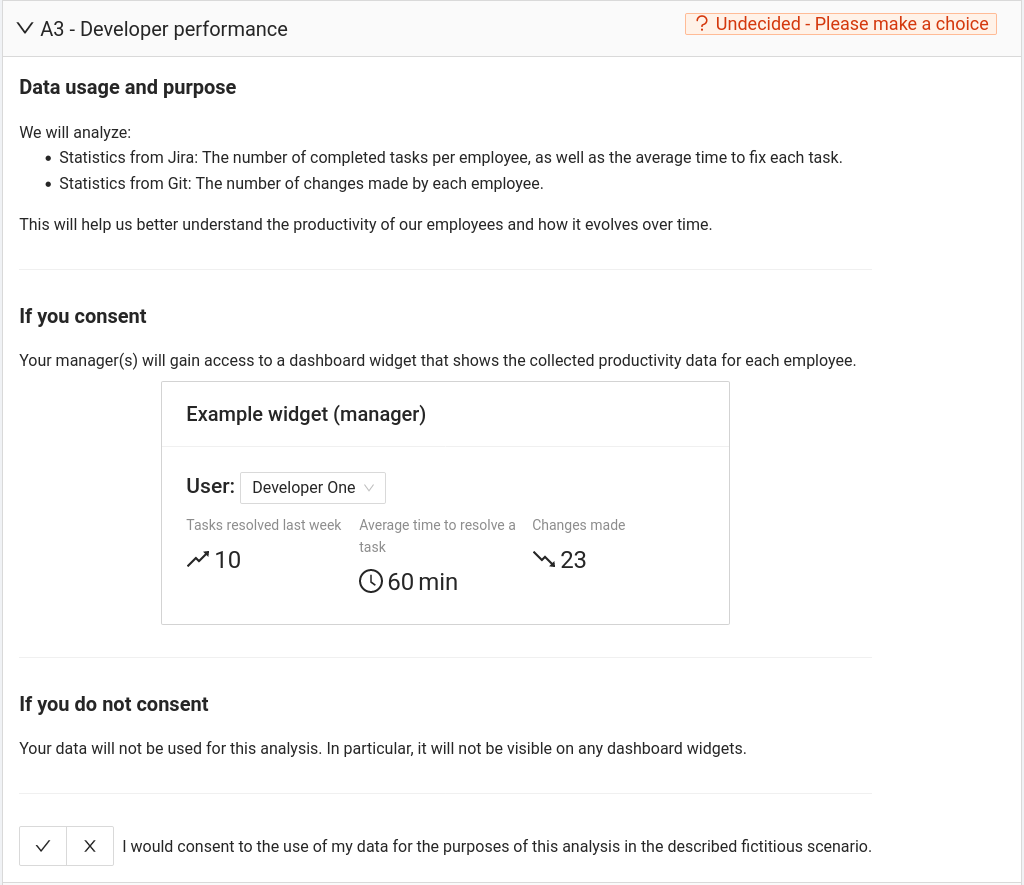}}
      \caption{Description of \analysisDP{} (developer performance) without data owner benefits in our exemplary implementation.}
      \Description{A screenshot of a website. At the top, a list entitled ``data usage and purpose,'' noting that statistics from Jira (number of completed tasks, average time to fix) and Git (number of changes made by each employee) will be analyzed. Below, a section ``If you consent,'' describing the analysis textually and presenting a screenshot. It is the analysis shown in Figure 6. Finally, a section ``If you do not consent,'' describing that data won't be used then. At the bottom, two buttons, one to confirm, one to reject sharing.}
      \label{fig:description-a3-base}
    \end{figure}

    \begin{figure}[H]
      \fbox{\includegraphics[width=.95\textwidth]{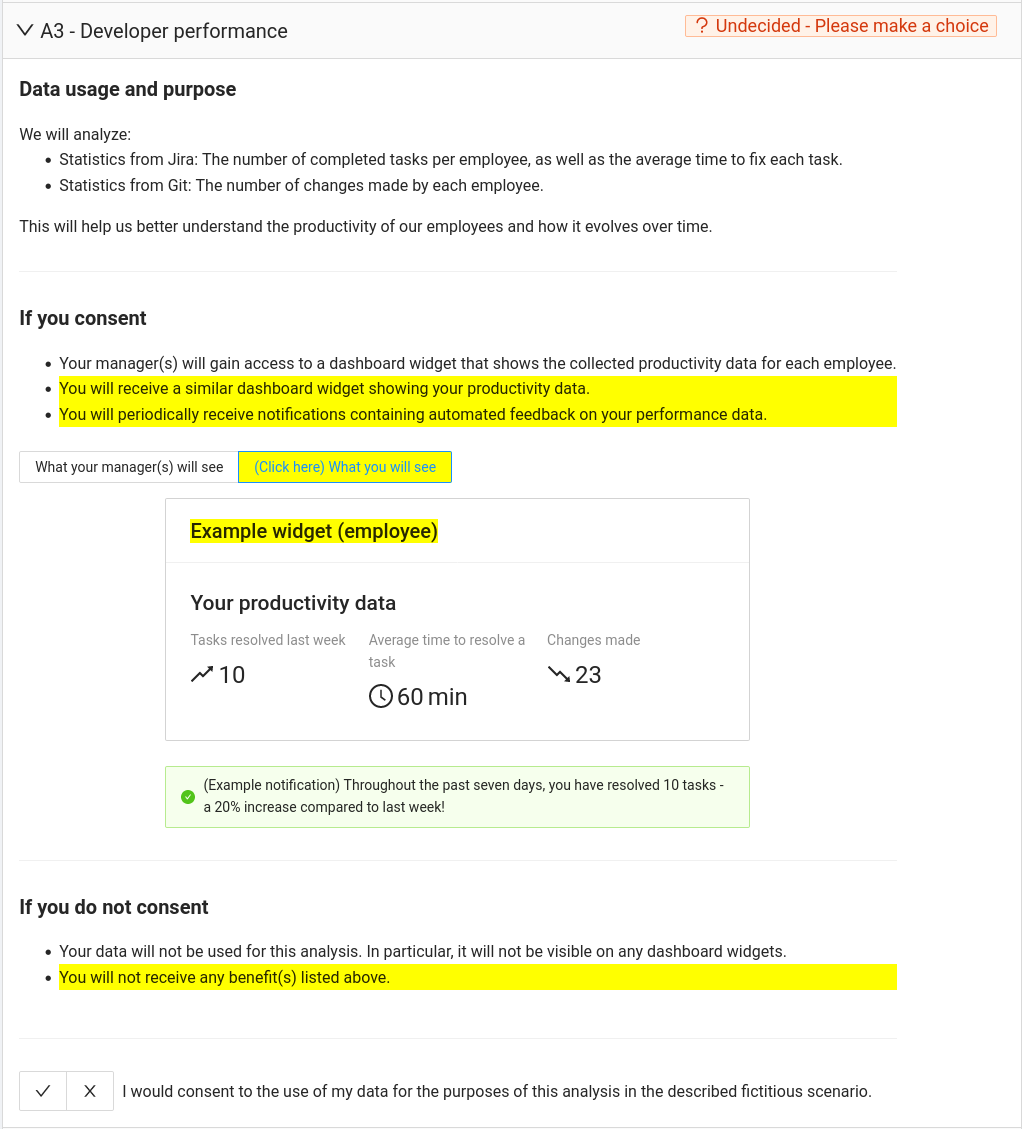}}
      \caption{Description of \analysisDP{} (developer performance) with data owner benefits in our exemplary implementation.}
      \Description{A screenshot of a website. It is the same as in Figure 16, with the only change being that now, the section ``If you consent'' additionally lists benefits, namely that the user will receive a similar widget as the manager, focusing on their own productivity data. The widget is otherwise identical.}
      \label{fig:description-a3-changed}
    \end{figure}

    \begin{figure}[H]
      \fbox{\includegraphics[width=.95\textwidth]{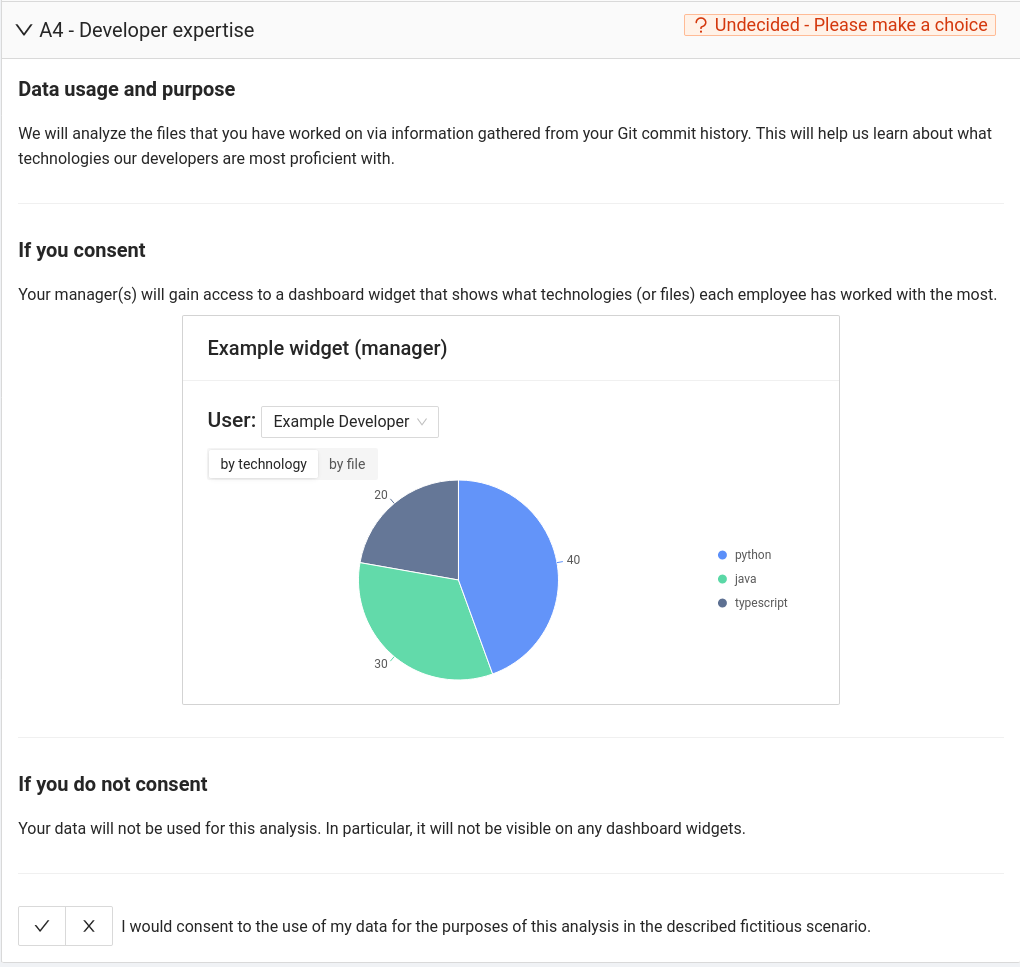}}
      \caption{Description of \analysisDE{} (developer expertise) without data owner benefits in our exemplary implementation.}
      \Description{A screenshot of a website. At the top, a list entitled ``data usage and purpose,'' noting that the files the employee worked on will be analyzed to determine the technologies that they employ. Below, a section ``If you consent,'' describing the analysis textually and presenting a screenshot. It is the analysis shown in Figure 7. Finally, a section ``If you do not consent,'' describing that data won't be used then. At the bottom, two buttons, one to confirm, one to reject sharing.}
      \label{fig:description-a4-base}
    \end{figure}

    \begin{figure}[H]
      \fbox{\includegraphics[width=.95\textwidth]{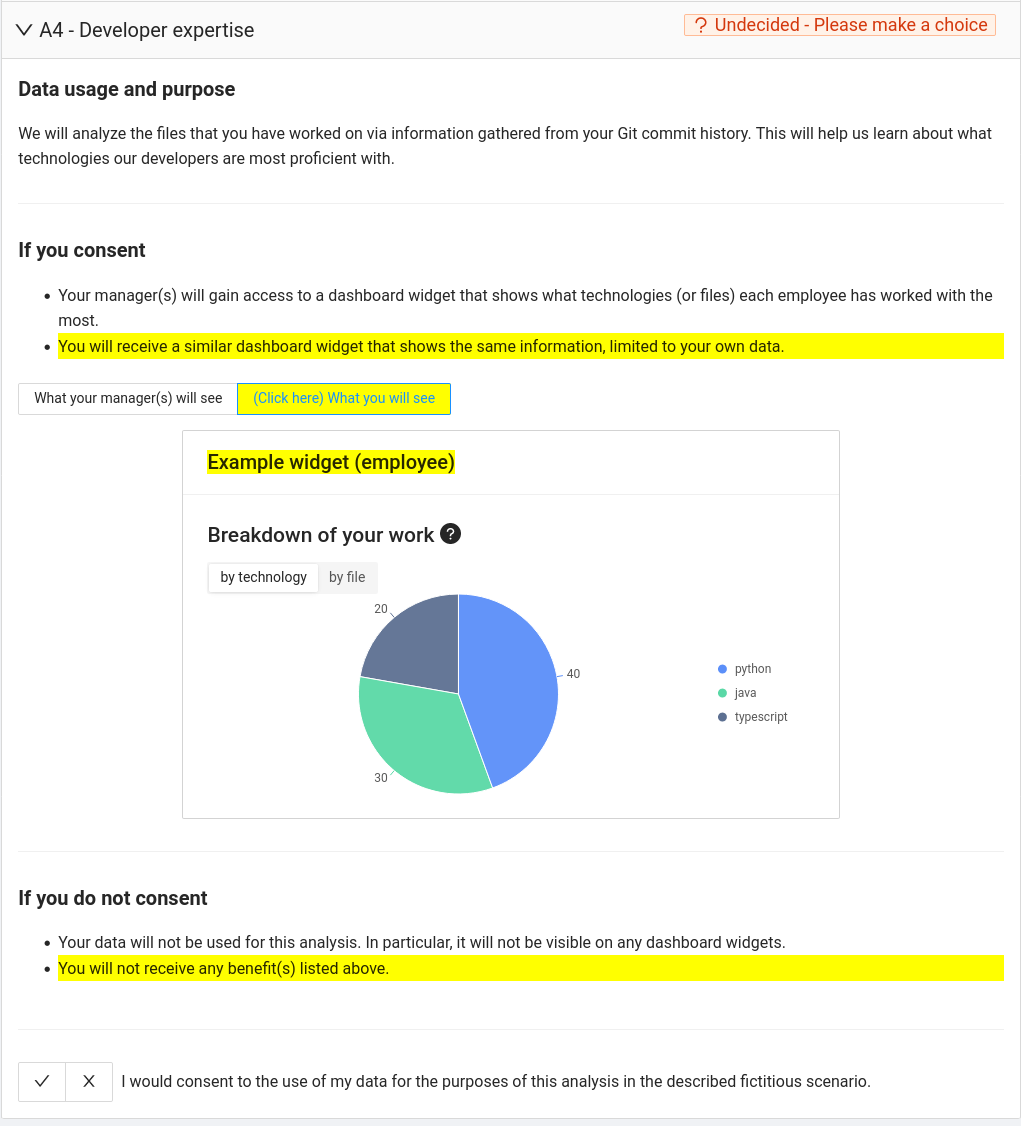}}
      \caption{Description of \analysisDE{} (developer expertise) with data owner benefits in our exemplary implementation.}
      \Description{A screenshot of a website. It is the same as in Figure 18, with the only change being that now, the section ``If you consent'' additionally lists benefits, namely that the user will receive insights about themselves. Accordingly, the widget shows the technologies that they worked with, instead of the whole team.}
      \label{fig:description-a4-changed}
    \end{figure}

    \received{October 2022}
    \received[revised]{February 2023}
    \received[accepted]{April 2023}
\end{document}